\documentclass[twocolumn]{aastex63}

\usepackage{amsmath}	
\usepackage{amssymb}	
\usepackage{enumitem}
\usepackage{verbatim} 
\usepackage[caption=false]{subfig}
\usepackage{array}

\newcommand{\bRcompanionRearth}{$1.586\pm0.098$} 
\newcommand{\baAU}{$0.0469\pm0.0017$} 
\newcommand{\bTeq}{$1099_{-18}^{+19}$} 
\newcommand{\bdepthtrundilutedTESS}{$0.302\pm0.031$} 
\newcommand{\bdepthtrdilutedTESS}{$0.302\pm0.031$} 
\newcommand{\cRcompanionRearth}{$2.068_{-0.091}^{+0.10}$} 
\newcommand{\caAU}{$0.0651\pm0.0024$} 
\newcommand{\cTeq}{$932_{-16}^{+17}$} 
\newcommand{\cdepthtrundilutedTESS}{$0.517_{-0.040}^{+0.036}$} 
\newcommand{\cdepthtrdilutedTESS}{$0.517_{-0.040}^{+0.036}$} 
\newcommand{\cperiodoverbperiod}{$1.63461_{-0.00025}^{+0.00026}$} 
\newcommand{\dRcompanionRearth}{$2.72\pm0.11$} 
\newcommand{\daAU}{$0.1131\pm0.0040$} 
\newcommand{\dTeq}{$708_{-12}^{+13}$} 
\newcommand{\ddepthtrundilutedTESS}{$0.889\pm0.053$} 
\newcommand{\ddepthtrdilutedTESS}{$0.889\pm0.053$} 
\newcommand{\dperiodovercperiod}{$2.28501\pm0.00027$} 
\newcommand{\eRcompanionRearth}{$3.12_{-0.12}^{+0.13}$} 
\newcommand{\eaAU}{$0.1400\pm0.0052$} 
\newcommand{\eTeq}{$636_{-11}^{+12}$} 
\newcommand{\edepthtrundilutedTESS}{$1.175\pm0.069$} 
\newcommand{\edepthtrdilutedTESS}{$1.175\pm0.069$} 
\newcommand{\eperiodoverdperiod}{$1.38208_{-0.00017}^{+0.00019}$} 

\newcommand{\bperiod}{$3.79523_{-0.00044}^{+0.00047}$} 
\newcommand{\cperiod}{$6.20370_{-0.00052}^{+0.00064}$} 
\newcommand{\dperiod}{$14.17555_{-0.0011}^{+0.00099}$} 
\newcommand{\eperiod}{$19.5917_{-0.0020}^{+0.0022}$} 

\newcommand{\rjup}{\mbox{R$_{J}$}}

\newcommand{\rearth}{R$_{\oplus}$}

\received{23 April 2020}
\accepted{29 December 2020}
\submitjournal{AJ}

\begin{document}

\title{TESS discovery of a super-Earth and three sub-Neptunes hosted by the bright, Sun-like star HD~108236\footnote{This paper includes data gathered with the 6.5 meter Magellan Telescopes located at Las Campanas Observatory, Chile.}}
\shorttitle{TESS discovery of the HD~108236 multiplanetary system}
\shortauthors{Daylan et al.}

\suppressAffiliations

\correspondingauthor{Tansu~Daylan}
\email{tdaylan@mit.edu}

\author[0000-0002-6939-9211]{Tansu Daylan}
\affil{Department of Physics and Kavli Institute for Astrophysics and Space Research, Massachusetts Institute of Technology, 70 Vassar Street, Cambridge, MA 02139, USA}
\affil{Kavli Fellow}

\author[0000-0001-8965-1472]{Kartik~Pingl\'e}
\affil{Cambridge Rindge and Latin School}

\author[0000-0002-0179-2105]{Jasmine~Wright}
\affil{Bedford High School}


\author[0000-0002-3164-9086]{Maximilian~N.~G{\"u}nther}
\affiliation{Department of Physics and Kavli Institute for Astrophysics and Space Research, Massachusetts Institute of Technology, 70 Vassar Street, Cambridge, MA 02139,
USA}
\affiliation{Juan Carlos Torres Fellow}

\author[0000-0002-3481-9052]{Keivan~G.~Stassun} 
\affiliation{Department of Physics and Astronomy, Vanderbilt University, Nashville, TN 37235, USA} 
\affiliation{Department of Physics, Fisk University, Nashville, TN 37208, USA}

\author[0000-0002-7084-0529]{Stephen~R.~Kane}
\affiliation{Department of Earth and Planetary Sciences, University of California, Riverside, CA 92521, USA}

\author[0000-0001-7246-5438]{Andrew~Vanderburg}
\affil{Department of Astronomy, The University of Texas at Austin, 2515 Speedway, Stop C1400, Austin, TX 78712, USA}
\affil{NASA Sagan Fellow}

\author[0000-0001-7246-5438]{Daniel~Jontof-Hutter}
\affil{Dept. of Physics, University of the Pacific, 3601 Pacific Avenue, Stockton, CA 95211, USA}

\author[0000-0001-8812-0565]{Joseph~E.~Rodriguez}
\affiliation{Center for Astrophysics \textbar Harvard \& Smithsonian, 60 Garden Street, Cambridge, MA 02138, USA}

\author[0000-0002-1836-3120]{Avi~Shporer}
\affil{Department of Physics and Kavli Institute for Astrophysics and Space Research, Massachusetts Institute of Technology, 70 Vassar Street, Cambridge, MA 02139, USA}

\author[0000-0003-0918-7484]{Chelsea~X.~Huang}
\affiliation{Department of Physics and Kavli Institute for Astrophysics and Space Research, Massachusetts Institute of Technology, 70 Vassar Street, Cambridge, MA 02139, USA}
\affil{Juan Carlos Torres Fellow}

\author[0000-0001-5442-1300]{Thomas~Mikal-Evans}
\affil{Department of Physics and Kavli Institute for Astrophysics and Space Research, Massachusetts Institute of Technology, 70 Vassar Street, Cambridge, MA 02139, USA}

\author[0000-0003-4903-567X]{Mariona~Badenas-Agusti}
\affil{Department of Earth, Atmospheric, and Planetary Sciences, Massachusetts Institute of Technology, Cambridge, MA 02139, USA}
\affil{Department of Physics and Kavli Institute for Astrophysics and Space Research, Massachusetts Institute of Technology, 70 Vassar Street, Cambridge, MA 02139, USA}

\author[0000-0001-6588-9574]{Karen~A.~Collins}
\affiliation{Center for Astrophysics \textbar Harvard \& Smithsonian, 60 Garden Street, Cambridge, MA 02138, USA}

\author[0000-0002-3627-1676]{Benjamin~V.~Rackham}
\affil{Department of Earth, Atmospheric and Planetary Sciences, and Kavli Institute for Astrophysics and Space Research, Massachusetts Institute of Technology, Cambridge, MA 02139, USA}
\affil{51 Pegasi b Fellow}

\author[0000-0002-8964-8377]{Samuel~N.~Quinn}
\affiliation{Center for Astrophysics \textbar Harvard \& Smithsonian, 60 Garden Street, Cambridge, MA 02138, USA}

\author[0000-0001-5383-9393]{Ryan~Cloutier}
\affiliation{Center for Astrophysics \textbar Harvard \& Smithsonian, 60 Garden Street, Cambridge, MA 02138, USA}

\author[0000-0003-2781-3207]{Kevin~I.~Collins}
\affiliation{George Mason University, 4400 University Drive, Fairfax, VA 22030 USA}

\author[0000-0002-4308-2339]{Pere~Guerra}
\affiliation{Observatori Astronòmic Albanyà, Camí de Bassegoda S/N, Albanyà 17733, Girona, Spain}

\author[0000-0002-4625-7333]{Eric~L.~N.~Jensen}
\affiliation{Dept. of Physics \& Astronomy, Swarthmore College, Swarthmore, PA 19081, USA}

\author[0000-0003-0497-2651]{John~F.~Kielkopf} 
\affiliation{Department of Physics and Astronomy, University of Louisville, Louisville, KY 40292, USA}

\author[0000-0001-8879-7138]{Bob~Massey}
\affiliation{Villa 39 Observatory, Landers, CA 92285, USA}

\author[0000-0001-8227-1020]{Richard~P.~Schwarz}
\affiliation{Patashnick Voorheesville Observatory, Voorheesville, NY 12186, USA}

\author[0000-0002-9003-484X]{David~Charbonneau}
\affiliation{Center for Astrophysics \textbar Harvard \& Smithsonian, 60 Garden Street, Cambridge, MA 02138, USA}

\author[0000-0001-6513-1659]{Jack~J.~Lissauer}
\affiliation{NASA Ames Research Center, Moffett Field, CA 94035, USA}

\author{Jonathan~M.~Irwin}
\affiliation{Center for Astrophysics \textbar Harvard \& Smithsonian, 60 Garden Street, Cambridge, MA 02138, USA}

\author[0000-0002-4746-0181]{{\"O}zg{\"u}r~Ba\c{s}t{\"u}rk}
\affil{Ankara University, Faculty of Science, Astronomy \& Space Sciences Dept. E Blok 205, TR-06100, Ankara, Turkey}

\author[0000-0003-3504-5316]{Benjamin~Fulton}
\affil{NASA Exoplanet Science Institute / Caltech-IPAC, 770 S. Wilson Avenue, Pasadena, CA 91125, USA}

\author[0000-0002-0345-2147]{Abderahmane~Soubkiou}
\affiliation{Oukaimeden Observatory, High Energy Physics and Astrophysics Laboratory, Cadi Ayyad University, Morocco}

\author[0000-0001-6285-9847]{Benkhaldoun~Zouhair}
\affiliation{Oukaimeden Observatory, High Energy Physics and Astrophysics Laboratory, Cadi Ayyad University, Morocco}

\author[0000-0002-2532-2853]{Steve~B.~Howell}
\affiliation{NASA Ames Research Center, Moffett Field, CA 94035, USA}

\author[0000-0002-0619-7639]{Carl~Ziegler}
\affiliation{Dunlap Institute for Astronomy and Astrophysics, University of Toronto, 50 St. George Street, Toronto, Ontario M5S 3H4, Canada}

\author[0000-0001-7124-4094]{C\'{e}sar~Brice\~{n}o}
\affiliation{Cerro Tololo Inter-American Observatory/NSF’s NOIRLab, Casilla 603, La Serena, Chile} 

\author[0000-0001-9380-6457]{Nicholas~Law}
\affiliation{Department of Physics and Astronomy, The University of North Carolina at Chapel Hill, Chapel Hill, NC 27599, USA}

\author[0000-0003-3654-1602]{Andrew W. Mann}
\affiliation{Department of Physics and Astronomy, The University of North Carolina at Chapel Hill, Chapel Hill, NC 27599, USA} 

\author{Nic~Scott}
\affiliation{NASA Ames Research Center, Moffett Field, CA 94035, USA}

\author[0000-0001-9800-6248]{Elise~Furlan}
\affil{NASA Exoplanet Science Institute / Caltech-IPAC, 770 S. Wilson Avenue, Pasadena, CA 91125, USA}

\author[0000-0002-5741-3047]{David~R.~Ciardi}
\affil{NASA Exoplanet Science Institute / Caltech-IPAC, 770 S. Wilson Avenue, Pasadena, CA 91125, USA}

\author[0000-0002-3321-4924]{Rachel~Matson}
\affil{U.S. Naval Observatory, Washington, DC 20392, USA}

\author[0000-0002-3439-1439]{Coel~Hellier}
\affil{Astrophysics Group, Keele University, Staffordshire ST5 5BG, United Kingdom}

\author[0000-0001-7416-7522]{David~R.~Anderson}
\affil{Astrophysics Group, Keele University, Staffordshire ST5 5BG, United Kingdom}
\affil{Department of Physics, University of Warwick, Gibbet Hill Road, Coventry CV4 7AL, UK}

\author[0000-0003-1305-3761]{R.~Paul~Butler}
\affil{Earth \& Planets Laboratory, Carnegie Institution for Science, 5241 Broad Branch Road, NW, Washington, DC 20015, USA}

\author[0000-0002-5226-787X]{Jeffrey~D.~Crane}
\affil{Observatories of the Carnegie Institution for Science, 813 Santa Barbara Street, Pasadena, CA 91101, USA}

\author{Johanna~K.~Teske}
\affil{Observatories of the Carnegie Institution for Science, 813 Santa Barbara Street, Pasadena, CA 91101, USA}
\affil{NASA fellow}

\author{Stephen~A.~Shectman}
\affil{Observatories of the Carnegie Institution for Science, 813 Santa Barbara Street, Pasadena, CA 91101, USA}


\author[0000-0002-2607-138X]{Martti~H.~Kristiansen}
\affil{Brorfelde Observatory, Observator Gyldenkernes Vej 7, DK-4340 T\o{}ll\o{}se, Denmark}
\affil{DTU Space, National Space Institute, Technical University of Denmark, Elektrovej 327, DK-2800 Lyngby, Denmark}

\author{Ivan~A.~Terentev}  
\affiliation{Citizen Scientist, Planet Hunter, Petrozavodsk, Russia}

\author{Hans~Martin~Schwengeler}
\affiliation{Citizen Scientist, Planet Hunter, Bottmingen, Switzerland}   

\author[0000-0003-2058-6662]{George~R.~Ricker}
\affil{Department of Physics and Kavli Institute for Astrophysics and Space Research, Massachusetts Institute of Technology, 70 Vassar Street, Cambridge, MA 02139, USA}

\author[0000-0001-6763-6562]{Roland~Vanderspek}
\affil{Department of Physics and Kavli Institute for Astrophysics and Space Research, Massachusetts Institute of Technology, 70 Vassar Street, Cambridge, MA 02139, USA}

\author[0000-0002-6892-6948]{Sara~Seager}
\affil{Department of Physics and Kavli Institute for Astrophysics and Space Research, Massachusetts Institute of Technology, 70 Vassar Street, Cambridge, MA 02139, USA}
\affil{Department of Earth, Atmospheric, and Planetary Sciences, Massachusetts Institute of Technology, Cambridge, MA 02139, USA}
\affiliation{Department of Aeronautics and Astronautics, MIT, 77 Massachusetts Avenue, Cambridge, MA 02139, USA}

\author[0000-0002-4265-047X]{Joshua~N.~Winn}
\affiliation{Department of Astrophysical Sciences, Princeton University, 4 Ivy Lane, Princeton, NJ 08544, USA}

\author[0000-0002-4715-9460]{Jon~M.~Jenkins}
\affiliation{NASA Ames Research Center, Moffett Field, CA 94035, USA}

\author[0000-0002-3321-4924]{Zachory~K.~Berta-Thompson} \affiliation{Department of Astrophysical and Planetary Science, University of Colorado Boulder, Boulder, CO 80309, USA}

\author[0000-0002-0514-5538]{Luke~G.~Bouma}
\affiliation{Department of Astrophysical Sciences, Princeton University, 4 Ivy Lane, Princeton, NJ 08544, USA}

\author[0000-0003-0241-2757]{William~Fong}
\affiliation{Department of Physics and Kavli Institute for Astrophysics and Space Research, Massachusetts Institute of Technology, 70 Vassar Street, Cambridge, MA 02139, USA}

\author{Gabor~Furesz}
\affiliation{Department of Physics and Kavli Institute for Astrophysics and Space Research, Massachusetts Institute of Technology, 70 Vassar Street, Cambridge, MA 02139, USA} 

\author{Christopher~E.~Henze}
\affiliation{NASA Ames Research Center, Moffett Field, CA 94035, USA}

\author[0000-0003-1447-6344]{Edward~H.~Morgan}
\affiliation{Department of Physics and Kavli Institute for Astrophysics and Space Research, Massachusetts Institute of Technology, 70 Vassar Street, Cambridge, MA 02139, USA}

\author[0000-0003-1309-2904]{Elisa~Quintana}
\affil{NASA Goddard Space Flight Center, 8800 Greenbelt Rd, Greenbelt, MD 20771}

\author[0000-0002-8219-9505]{Eric~B.~Ting}
\affiliation{NASA Ames Research Center, Moffett Field, CA 94035, USA}

\author[0000-0002-6778-7552]{Joseph~D.~Twicken}
\affiliation{NASA Ames Research Center, Moffett Field, CA 94035, USA} 
\affiliation{SETI Institute, 189 Bernardo Ave, Suite 100, Mountain View, CA 94043, USA}









\begin{abstract}

We report the discovery and validation of four extrasolar planets hosted by the nearby, bright, Sun-like (G3V) star HD~108236 using data from the Transiting Exoplanet Survey Satellite (TESS). We present transit photometry, reconnaissance and precise Doppler spectroscopy as well as high-resolution imaging, to validate the planetary nature of the objects transiting HD~108236, also known as the TESS Object of Interest (TOI) 1233. The innermost planet is a possibly-rocky super-Earth with a period of \bperiod{} days and has a radius of \bRcompanionRearth{} $R_\oplus$. The outer planets are sub-Neptunes, with potential gaseous envelopes, having radii of \cRcompanionRearth{} $R_\oplus$, \dRcompanionRearth{} $R_\oplus$, and \eRcompanionRearth{} $R_\oplus$ and periods of \cperiod{} days, \dperiod{} days, and \eperiod{} days, respectively. With V and K$_{\rm s}$ magnitudes of 9.2 and 7.6, respectively, the bright host star makes the transiting planets favorable targets for mass measurements and, potentially, for atmospheric characterization via transmission spectroscopy. HD~108236 is the brightest Sun-like star in the visual (V) band known to host four or more transiting exoplanets. The discovered planets span a broad range of planetary radii and equilibrium temperatures, and share a common history of insolation from a Sun-like star ($R_\star = 0.888 \pm 0.017$\,R$_\odot$, $T_{\rm eff} = 5730 \pm 50$~K), making HD~108236 an exciting, opportune cosmic laboratory for testing models of planet formation and evolution.
\end{abstract}


\section{Introduction}
\label{sect:intr}

As the number and diversity of the known exoplanets continues to grow, we are gaining a better perspective on our own Solar System. Based on the discovery of more than 4,000 exoplanets\footnote{https://exoplanetarchive.ipac.caltech.edu} to date \citep{Akeson+2013}, two common types of exoplanets are the larger analogs of the Earth (super-Earths)\footnote{Throughout this paper, we refer to a planet as a super-Earth or sub-Neptune if its radius is smaller than 1.8$R_{\Earth}$ and between 1.8$R_{\Earth}$ and 4$R_{\Earth}$, respectively} and smaller analogs of Neptune (sub-Neptunes) \citep{Fressin+2013, Fulton+2017}. Their wide range of orbital architectures and atmospheric properties \citep{Kite+2020, Rein2012} motivate further investigation of these small exoplanets in order to accurately characterize their demographic properties.

Transiting exoplanets hosted by bright stars enable detailed characterization such as measurements of radius, mass, bulk composition and atmospheric properties. Furthermore, multiplanetary systems offer laboratories to study how planet formation, evolution and habitability depend on amount of insolation, while controlling for the age and stellar type \citep{PuWu2015, Weiss+2018a, Weiss+2018b}.

The Transiting Exoplanet Survey Satellite (TESS) \citep{Ricker+2014} is a spaceborne NASA mission launched in 2018 to survey the sky for transiting exoplanets around nearby and bright stars. It builds on the legacy of the NASA's Kepler space telescope \citep{Borucki+2010}
launched in 2009, which was the first exoplanet mission to perform a large statistical survey of transiting exoplanets. One of the goals of the TESS mission is to discover 50 exoplanets with radii smaller than 4$R_{\Earth}$ and coordinate their mass measurements via precise high-resolution spectroscopic follow-up. This will enable accurate inferences about the bulk composition and atmospheric characterization of small exoplanets.

In this work, we present the discovery and validation of four exoplanets hosted by HD~108236, also identified as the TESS Object of Interest (TOI) 1233. We use the TESS data in sectors 10 and 11 (i.e., UT 26 March 2019 to UT 21 May 2019) as well as ground-based follow-up data to validate the planetary nature of the transits detected in the TESS data and precisely determine the properties of the planets and their host star.

HD~108236 is the brightest Sun-like (G-type) star and one of the brightest stars on the sky to host at least four transiting planets. This makes it an especially useful system for comparative studies of the formation and evolution of its transiting planets in the future. Furthermore, its planets are favorable targets for atmospheric characterization via transmission spectroscopy. With a super-Earth and three sub-Neptunes, the HD~108236 system constitutes a major contribution to the mission goal of TESS. HD~108236 is also the first multiplanetary system delivered by TESS with four validated transiting planets.

This paper is organized as follows. In Section \ref{sect:star}, we characterize the host star HD~108236. In Section \ref{sect:vali}, we present the data collected on the system to discover and validate the planets. We then characterize the planets in Section \ref{sect:plan}, discuss our results and conclude in Section \ref{sect:disc}.

\section{Stellar characterization}
\label{sect:star}

Characterization of an exoplanet, i.e., determination of its mass, $M_{\rm p}$, radius, $R_{\rm p}$, and equilibrium temperature, $T_{\rm eq}$, requires determination of the same properties of its host star. Therefore, we first study and characterize the host star to estimate its radius, $R_\star$, mass, $M_\star$, and effective temperature, $T_{\rm eff}$, as well as its surface gravity, $\log g$, metallicity, [Fe/H], sky-projected rotational velocity, $v \sin i_\star$, and spectroscopic class.

HD~108236 is a bright main-sequence star with a TESS magnitude of 8.65 in the Southern Ecliptic Hemisphere, falling in the Centaurus constellation with a right ascension and declination of 12:26:17.78 -51:21:46.99 (186.574063$^\circ$ -51.363052$^\circ$). Having a parallax of $15.45 \pm 0.05$ milli arcsecond (mas) as measured by the Gaia telescope in its Data Release 2 (DR2) \citep{Brown+2018, Bailer-Jones+2018}, the host star is 64.6 $\pm$ 0.2 parsecs away. Based on the same Gaia DR2 catalog, it has a proper motion of $-70.43 \pm 0.06$ and $-49.87 \pm 0.04$ mas per year along right ascension and declination, respectively, and a velocity along our line of sight of $16.78\pm0.02$\,km s$^{-1}$. Although we will be referring to the star as HD~108236 throughout this work, some other designations for the target are TIC 260647166, TOI~1233, and HIP 60689.

\begin{table}[]
    \centering
    \caption{Stellar Information}
    \begin{flushleft}
    \underline{Identifying Information}
    \end{flushleft}

    \begin{tabular}{m{2.65cm} m{3.1cm} m{1.75cm}}
        Name & TOI~1233, HD~108236 & \\ 
        TIC ID & 260647166 & \\
        \\
    \end{tabular}\\

    \begin{tabular}{m{2.65cm} | m{3.1cm} | m{1.75cm}}
        \hline
        Parameter & Value & Reference\\
        \hline
    \end{tabular}\\

    \begin{flushleft}
    \underline{Astrometric Properties}
    \end{flushleft}

    \begin{tabular}{m{2.65cm} | m{3.1cm} | m{1.75cm}}
        \text{Right Ascension [$^{\circ}$]} & 186.574063 & Gaia DR2\\
        \text{Declination [$^{\circ}$]} & -51.363052 & Gaia DR2\\
        \text{$\mu_{\alpha}$} \text{[mas yr$^{-1}$]} & -70.43 $\pm$ 0.06 & Gaia DR2\\
        \text{$\mu_{\delta}$} \text{[mas yr$^{-1}$]} & -49.87 $\pm$ 0.04 & Gaia DR2\\
        Distance [pc] & 64.6 $\pm$ 0.2 & TIC v8\\
        RV [km s$^{-1}$] & $16.78\pm0.02$ km s$^{-1}$ & Gaia DR2 \\
    \end{tabular}\\
    
    \begin{flushleft}
    \underline{Photometric Properties}
    \end{flushleft}

    \begin{tabular}{m{2.65cm} | m{3.1cm} | m{1.75cm}}
        TESS [mag] & 8.6522  $\pm$  0.006 & TIC v8\\ 
        B [mag] & 9.89  $\pm$  0.02 & TIC v8\\   
        V [mag] & 9.25  $\pm$  0.01 & TIC v8\\   
        B$_{\rm T}$ [mag] & 10.04 $\pm$ 0.02 & Tycho-2\\
        V$_{\rm T}$ [mag] & 9.313 $\pm$ 0.014 & Tycho-2\\
        Gaia [mag] & 9.08745  $\pm$  0.0002 & Gaia DR2\\ 
        Gaia$_{\text{BP}}$ [mag] & 9.43555 $\pm$ 0.000737 & Gaia DR2\\ 
        Gaia$_{\text{RP}}$ [mag] & 8.60563 $\pm$ 0.000643 & Gaia DR2\\ 
        J [mag] & 8.046  $\pm$  0.024 & 2MASS\\   
        H [mag] & 7.703  $\pm$  0.029 & 2MASS\\
        K$_{\text{s}}$ [mag] & 7.637  $\pm$ 0.031 & 2MASS\\   
        \text{WISE 3.4 [mag]} & 7.613  $\pm$  0.029  & WISE\\
        \text{WISE 4.6 [mag]} & 7.673  $\pm$  0.021  & WISE\\
        \text{WISE 12 [mag]} & 7.638  $\pm$  0.017  & WISE\\
        \text{WISE 22 [mag]} & 7.51  $\pm$  0.098  & WISE\\
        \hline
    \end{tabular}
    
    \begin{minipage}{8cm} 
    In the table, mas stands for milli arcseconds. We use the following references: TESS Input Catalog version 8 (TICv8) \citep{Stassun+2019}, Gaia DR2 \citep{Brown+2018}, Tycho-2 \citep{Hog+2000}, 2MASS \citep{Cutri+2003}, and WISE \citep{Wright+2010}.
    \end{minipage}
    \label{tabl:parastar}
\end{table}

Since photometric transit observations only probe the planet-to-star radius ratio, the stellar radius needs to be determined precisely in order to infer the radii of the transiting planets. The stellar radius can be inferred using two independent methods. First, a high-resolution spectrum of the star can be used to derive the stellar parameters, by fitting it with a model spectrum obtained by linearly interpolating a library of template spectra \citep{Coelho+2005}. The resulting effective temperature and the distance to the star then yield the stellar radius via the Stefan-Boltzmann law. We used this method to characterize the star based on the high-resolution spectrum described in Section \ref{sect:lcocnres}, obtaining the stellar radius and effective temperature as $0.894\pm0.022\,R_\odot$ and $5618\pm100$\,K, respectively.

\begin{figure}[!htbp]
    \centering
    \includegraphics[trim=0.5cm 0.5cm 0.5cm 0.5cm, clip, width=0.49\textwidth]{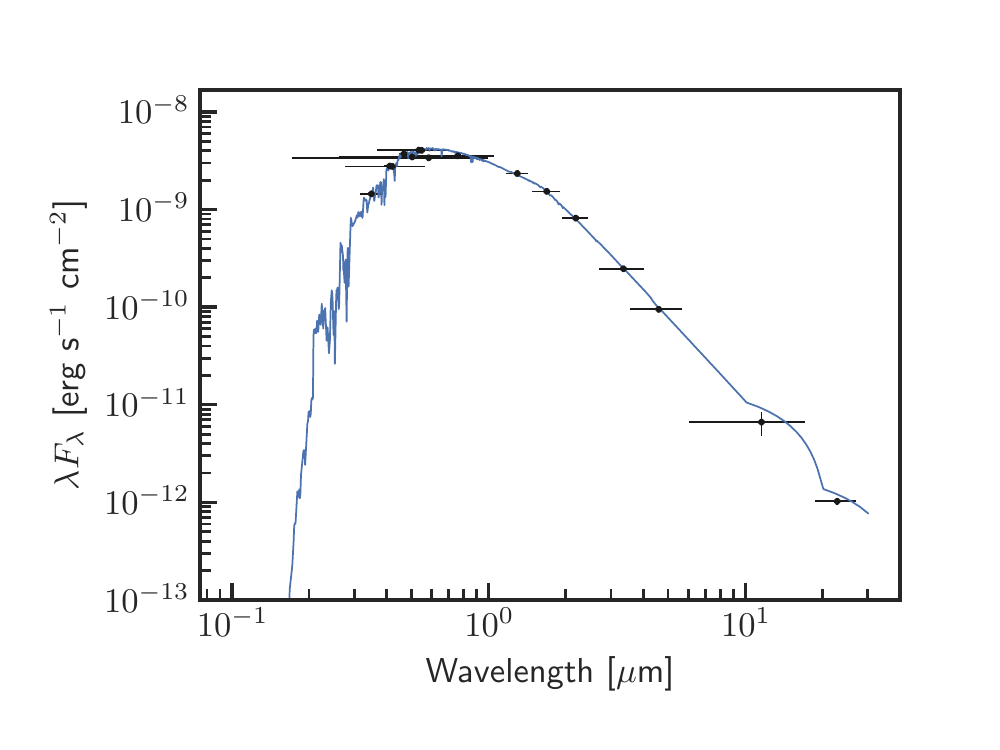}
    \caption{The SED of HD~108236. Black symbols and their vertical error bars represent the photometric measurements that were previously available on the system. The horizontal bars represent the effective width of the passband. Overplotted with the blue line is our best-fit Kurucz atmosphere model, allowing us to characterize the star.}
    \label{figr:specstar}
\end{figure}

An independent method of inferring the effective temperature and radius of the host star is to model its brightness across broad bands over a larger wavelength range, known as the spectral energy distribution (SED). This yields a semi-empirical determination of the stellar radius as well as independent constraints on stellar evolution model parameters such as the stellar mass, metallicity and age. Towards this purpose, we used the broad-band photometric magnitudes of HD~108236 provided in Table~\ref{tabl:parastar} to model the stellar SED of HD~108236 following the methodology described in \citet{StassunTorres2016, Stassun+2017a, Stassun+2018}. To constrain the distance to the star, we used the Gaia DR2 parallaxes, adjusted by 82\,$\mu$as to account for the systematic offset reported by \citet[][]{StassunTorres2018}. We retrieved the $B_{\rm T}$ and $V_{\rm T}$ magnitudes from Tycho-2, the Str\"{o}mgren $ubvy$ magnitudes from \citet{Paunzen2015}, the $JHK_{\rm S}$ magnitudes from 2MASS \citep{Skrutskie2006, Cutri+2003}, the W1, W2, W3, and W4 magnitudes from WISE \citep{Wright+2010}, and the $G$, $G_{\rm BP}$, and $G_{\rm RP}$ magnitudes from Gaia \citep{Brown+2018, Bailer-Jones+2018}. Together, the available photometry spans the full stellar SED over the wavelength range 0.35-22~$\mu$m as shown Figure~\ref{figr:specstar}.  

We performed a fit using Kurucz stellar atmosphere models \citep{CastelliKurucz2003}, with the effective temperature, $T_{\rm eff}$, metallicity, [Fe/H], and surface gravity, $\log g$, adopted from the TIC \citep{Stassun+2019} as initial guesses. The only additional free parameter was the extinction ($A_{\rm V}$), which we restricted to be less than or equal to the maximum line-of-sight value from the dust maps of \citet{Schlegel+1998}. The resulting fit is excellent (Figure~\ref{figr:specstar}) with a reduced $\chi^2$ of 2.3 and best-fit $A_{\rm V} = 0.04 \pm 0.04$, $T_{\rm eff} = 5730 \pm 50$~K, $\log g = 4.5 \pm 0.5$, and [Fe/H] = $-0.3 \pm 0.5$. Integrating the (unreddened) model SED gives the bolometric flux at Earth, $F_{\rm bol} = 5.881 \pm 0.068 \times 10^{-9}$\, erg~s$^{-1}$~cm$^{-2}$. Taking the $F_{\rm bol}$ and $T_{\rm eff}$ together with the Gaia DR2 parallax gives the stellar radius, $R_\star = 0.888 \pm 0.017\,R_\odot$. Finally, we can use the empirical relations of \citet{Torres+2010} and a 6\% error from the empirical relation itself to estimate the stellar mass, $M_\star = 0.97 \pm 0.06\,M_\odot$; this, in turn, together with the stellar radius provides an empirical estimate of the mean stellar density, $\rho_\star = 1.94 \pm 0.16$\,g~cm$^{-3}$. Based on these properties, the spectral type of HD~108236 can be assigned as G3V \citep{PecautMamajek2013}.

In an alternative, isochrone-dependent approach, we also used \texttt{EXOFASTv2} \citep{Eastman+2019} to constrain the stellar parameters. We relied on the observed SED and the MESA isochrones and stellar tracks \citep{Dotter+2016, Choi+2016}. This approach forces the inference to match a theoretical star based on stellar evolution models. We imposed Gaussian priors on the Gaia DR2 parallax. We added 82\,$\mu$as to the reported value and 33\,$\mu$as in quadrature to the reported error, following the recommendation of \citet{StassunTorres2018}. We also imposed an upper limit on the extinction of 0.65 using the dust map of \citet{SchlaflyFinkbeiner2011}. In addition, we applied Gaussian priors on $T_{\rm eff}$ and [Fe/H] from the analysis of the high-resolution spectrum described in Section \ref{sect:lcocnres}.

The derived stellar parameters from all approaches are summarized in Table~\ref{tabl:starchar}. When characterizing the transiting planets in the remaining of this paper, we use the stellar radius and the effective temperature of $0.888\pm0.017$\,R$_\Sun$ and $5730\pm50$\,K, as inferred from the isochrone-independent approach based on the SED.

\begin{table}[]
    \centering
    \caption{Stellar Characterization. Different methods yield consistent models of the host star. The difference between our adopted stellar parameters (i.e., based on an isochrone-independent model of the broad-band photometry) and those of the EXOFAST results can be attributed to the tight isochrone priors of the latter. The difference with the NRES results is largely due to the differences in the information content of broad-band photometry and high-resolution spectra.}
    \begin{tabular}{m{3.5cm} | m{3.5cm}}
    \hline
    \hline
    Parameter & Value \\
    \hline
    \end{tabular}
    \underline{High-resolution spectroscopy} \\
    \begin{flushleft}
    CHIRON
    \end{flushleft}
    \begin{tabular}{m{3.5cm} | m{3.5cm}}
    $T_{\text{eff}} \text{ [K]}$ & 5638 \\
    log [g] & 4.39\\
    {[Fe/H]} & -0.22 \\
    $v$sin$i$ [km s$^{-1}$] & $<$4.7 (95\% CL) \\
    \end{tabular}
    \begin{flushleft}
    LCO/NRES
    \end{flushleft}
    \begin{tabular}{m{3.5cm} | m{3.5cm}}
    $T_{\text{eff}} \text{ [K]}$ & 5618 \text{$\pm$} 100  \\
    log [g] & 4.6 \text{$\pm$} 0.1  \\
    {[Fe/H]} & -0.26 \text{$\pm$} 0.06  \\
    $v$sin$i$ [km s$^{-1}$] & $<$ 2 (95\% CL) \\ 
    M$_*$ \text{[M}$_\Sun$] & 0.853 \text{$\pm$} 0.047  \\  
    R$_*$ \text{[R}$_\Sun$] & 0.894 \text{$\pm$} 0.022  \\ 
    \end{tabular}
    \underline{Broad-band photometry} \\
    \begin{flushleft}
    Isochrone-independent
    \end{flushleft}
    \begin{tabular}{m{3.5cm} | m{3.5cm}}
    $T_{\text{eff}} \text{ [K]}$ & 5730 \text{$\pm$} 50  \\
    log [g] & 4.5 \text{$\pm$} 0.5  \\
    {[Fe/H]} & -0.3 \text{$\pm$} 0.5  \\
    $A_{\text{v}}$ & 0.04 \text{$\pm$} 0.04\\
    $F_{\text{bol}}$ \text{[erg s$^{-1}$ cm$^{-2}$]} & 5.881 \text{$\pm$} 0.068 \text{$\times$} 10$^{-9}$\\
    M$_*$ \text{[M}$_\Sun$] & 0.97 \text{$\pm$} 0.06  \\  
    R$_*$ \text{[R}$_\Sun$] & 0.888 \text{$\pm$} 0.017  \\ 
    $\rho_*$ [g cm$^{-3}$] & 1.94 \text{$\pm$} 0.16\\
    \end{tabular}
    \begin{flushleft}
    Isochrone-dependent approach via \texttt{EXOFASTv2}
    \end{flushleft}
    \begin{tabular}{m{3.5cm} | m{3.5cm}}
    $T_{\text{eff}} \text{ [K]}$ & 5721 \text{$\pm$} 60  \\
    log [g] & 4.492 \text{$\pm$} 0.032  \\
    {[Fe/H]} & -0.253 \text{$\pm$} 0.062  \\
    Age \text{Gyr} & 5.8 \text{$\pm$} 4.1 \\
    $A_{\text{v}}$ & 0.04 \text{$\pm$} 0.04\\
    $L_*$ \text{[L}$_\Sun$] &  0.747 \text{$\pm$} 0.03 \\
    M$_*$ \text{[M}$_\Sun$] & 0.877 \text{$\pm$} 0.05  \\  
    R$_*$ \text{[R}$_\Sun$] & 0.88 \text{$\pm$} 0.017  \\ 
    $\rho_*$ [g cm$^{-3}$] & 1.82 \text{$\pm$} 0.15 \\
    \hline
    \end{tabular}
    \begin{minipage}{8cm}
        CL stands for confidence level.
    \end{minipage}
    \label{tabl:starchar}
\end{table}

\section{Discovery and validation of planets hosted by HD~108236}
\label{sect:vali}

In this section, we will describe the detection of transit signals consistent with transiting planets hosted by HD~108236 and the follow-up data we collected to rule out alternative hypotheses. Table~\ref{tabl:datatfop} summarizes the observations we carried out using the resources of the TESS Follow-up Observing Program (TFOP) to validate the planetary origin of the transits and characterize the planets and their host star. The subgroups of TFOP involved in this program were ground-based photometry (SG1), reconnaissance spectroscopy (SG2), high-resolution imaging (SG3), and precise Doppler spectroscopy (SG4).

\begin{table}[]
    \centering
    \caption{Observations conducted as part of the follow-up of HD~108236 after the detection of transits by TESS.}
    \begin{tabular}{c|c }
        \hline
        Date & Telescope/Instrument\\

        \hline
        \underline{Imaging} & \\
    
        2020-01-14 & Gemini/Zorro \\
        2020-03-12 & \\
        \hline
        2020-01-07 & SOAR/HRCam \\

        \hline
        \underline{Reconnaissance Spectroscopy} & \\
        
        2020-01-28 & \\
        2020-01-24 & \\
        2019-08-03 & SMARTS/CHIRON \\
        2019-07-04 & \\
        2019-07-02 & \\
        \hline
        2019-06-12 & LCOGT/NRES \\
        2019-06-23 & \\

        \hline
        \underline{Precise Doppler spectroscopy} & \\
        
        2019-07-12 & \\
        2019-07-15 & \\
        2019-07-16 & \\
        2019-07-18 & \\
        2019-07-20 & \\
        2019-08-08 & Magellan II/PFS\\
        2019-08-09 & \\
        2019-08-11 & \\
        2019-08-13 & \\
        2019-08-17 & \\
        2019-08-20 & \\
        2019-08-21 & \\
        \hline
    \end{tabular}
    
    \begin{tabular}{c|c|c|c}
        Photometric & & & \\
        \hline
        Date & Telescope & Instrument & TOI \\
        2020-03-17 & LCOGT-CTIO & Sinistro & 1233.01* \\
        2020-03-17 & MEarth-South & Apogee & 1233.01 \\
        2020-03-11 & LCOGT-CTIO & Sinistro  & 1233.03 \\
        2020-03-11 & LCOGT-CTIO & Sinistro & 1233.02 \\
        2020-03-11 & MEarth-South & Apogee & 1233.02 \\
        2020-03-03 & MEarth-South & Apogee & 1233.01 \\
        2020-03-02 & LCOGT-SAAO & Sinistro & 1233.01 \\
        2020-02-02 & LCOGT-SAAO & Sinistro & 1233.02 \\
        2020-01-31 & LCOGT-SAAO & Sinistro & 1233.03 \\
        2020-01-11 & LCOGT-SAAO & Sinistro & 1233.04 \\
        2020-01-11 & LCOGT-CTIO & Sinistro & 1233.02 \\
        \hline
    \end{tabular}

    \begin{minipage}{8cm}
        A * in the last column denotes a tentative detection of a transit on target.
    \end{minipage}
    \label{tabl:datatfop}
\end{table}

\subsection{TESS}

TESS is a spaceborne telescope with four cameras, each with four Charge-Coupled Devices (CCDs) with the primary mission of discovering small planets hosted by bright stars, enabled by its high-precision photometric capability in space \citep{Ricker+2014}. The Science Processing Operations Center (SPOC) pipeline \citep{Jenkins+2016} regularly calibrates and reduces TESS data, delivering Simple Aperture Photometry (SAP) \citep{Twicken+2010, Morris+2017} light curves as well as Presearch Data Conditioning (PDC) \citep{Stumpe+2012, Smith+2012, Stumpe+2014} light curves that are corrected for systematics. Then, it searches for periodic transits in the resulting light curves using the Transiting Planet Search (TPS) \citep{Jenkins2002, Jenkins+2017} to search for planets. Unlike the Box Least Squares (BLS) \citep{Kovacs+2002}, which also searches for transit-like pulse trains while not taking into account the correlation structure of noise, TPS employs a noise-compensating matched filter which jointly characterizes the correlation structure of the observation noise while searching for periodic transits. Finally, it delivers the statistically significant candidates as Threshold Crossing Events (TCEs). As members of the TOI working group, we regularly classify these TCEs as planet candidates and false positives. When vetting TCEs as planet candidates, we use the SPOC validation tests \citep{Twicken+2018, Li+2019} such as:

\begin{enumerate}
\item the eclipsing binary discrimination test to detect the presence of secondary eclipses and compare the depths of odd and even transits to rule out inconsistencies,
\item the centroid offset test to determine if the centroid of the difference (i.e., out-of-transit minus in-transit) image is statistically consistent with the location of the target star,
\item a statistical bootstrap test to estimate the false positive probability of the TCE when compared to other transit-like features in the light curve, and
\item an optical ghost diagnostic test to rule out false positive hypotheses such as instrumental noise, scattered or blended light, based on the correlations between the model transit and light curves derived from the core photometric aperture and a surrounding halo.
\end{enumerate}

\subsection{Discovery of periodic transits consistent with planetary origin}

HD~108236 was among the list of targets observed by TESS with a cadence of 2 minutes and also included in the TESS Guest Investigator (GI) Cycle I proposal (G011250, PI: Walter, Frederick). It was observed by TESS Camera 2, CCD 2 during Sector 10 (UT 26 March 2019 - 22 April 2019) and TESS Camera 1, CCD1 during Sector 11 (UT 22 April 2019 - 21 May 2019). The TESS data were processed by the SPOC pipeline. Then, Sector 10 and 11 TESS data and derived products such as the SAP and PDC light curves including that of HD~108236 were made public on 01 June 2019 (data release 14) and 17 June 2019 (data release 16), respectively. 

\begin{figure*}[!htbp]
    \centering
    \includegraphics[trim=1cm 0cm 1cm 0.5cm, clip, width=\textwidth]{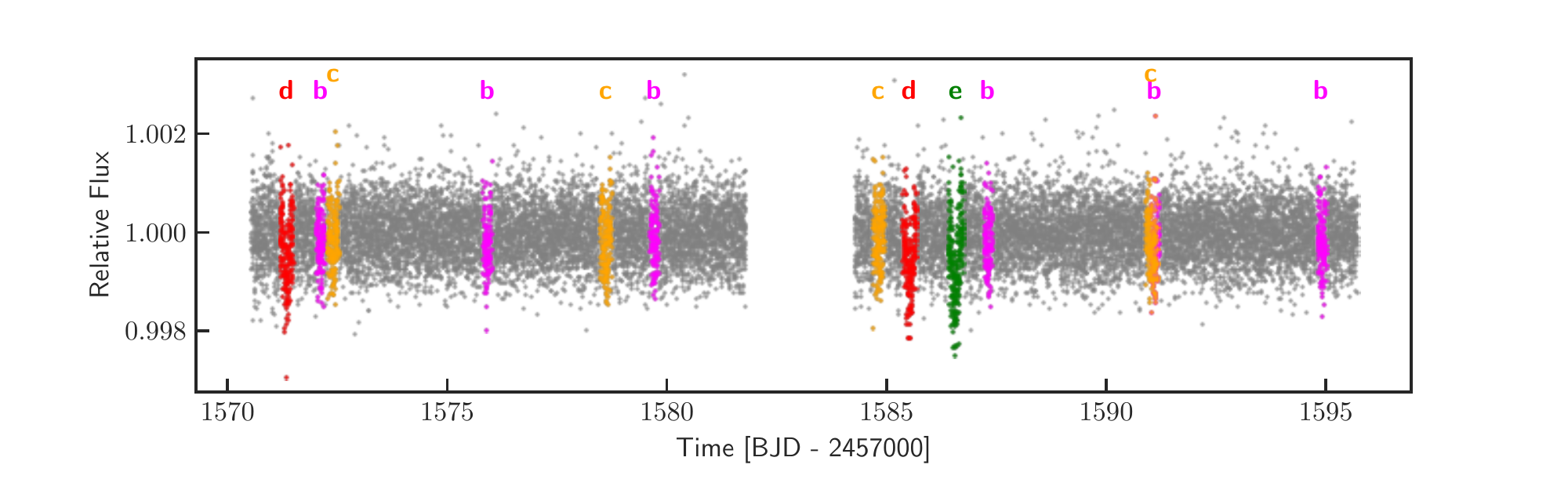} \\
    \includegraphics[trim=1cm 0cm 1cm 0.5cm, clip, width=\textwidth]{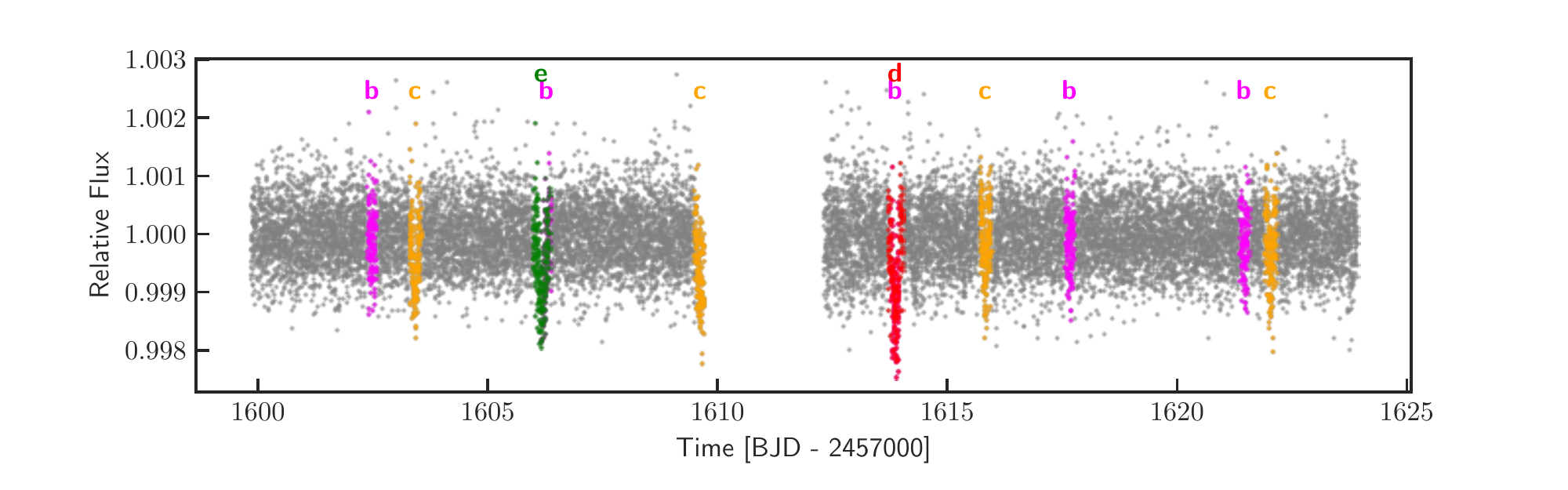}
    \caption{The normalized light curve of HD~108236 measured by TESS, reduced by the PDC pipeline, and detrended by our pipeline, shown with gray points. The top and bottom panels show the Sector 10 and 11 data, respectively. The data show stellar variability, especially in Sector 11, which is taken into account for both Sectors by our red noise model as discussed in Section~\ref{sect:modl}. Magenta, orange, red and green colors highlight the transits of the discovered planets b, c, d, and e. Throughout the paper we use the same color scheme to denote the planets.}
    \label{figr:lcur}
\end{figure*}

The first detection of a TCE consistent with a planetary origin from HD~108236 was obtained in Sector 10 TESS data. The TCE had a period of 14.178 days. However, the light curve also had other transit-like features unrelated to the detected TCE, which promoted HD~108236 to a potentially high-priority, multiplanetary system candidate. Sector 11 TESS data triggered three TCEs one of which had the same period as that from Sector 10. However, the transits of the other TCEs had inconsistent depths. These initial TCEs from individual sectors were vetted as planet candidates with the expectation that a joint TPS analysis of two sectors of TESS data would resolve the ambiguities on the multiplicity and periods of the planet candidates. The multi-sector data analysis at the end of Sector 13 resulted in the detection of four TCEs with periods 14.18, 19.59, 6.20, and 3.80 days and Signal to Noise Ratios (SNRs) 15.3, 16.2, 11.4, and 8.7, respectively. The PDC light curve of HD~108236 from these two sectors is shown in Figure~\ref{figr:lcur}. Subsequently, we released alerts on these four TCEs (i.e., TOI~1233.01, TOI~1233.02, TOI~1233.03, and TOI~1233.04) with planet candidate dispositions on 26 August 2019. For the moment, we will refer to these TCEs that have been vetted as planet candidates using the TOI designations.

\subsection{Vetting of the planet candidates}

Time-series photometry of a source is inferred from photoelectrons counted in a grid of pixels on the focal plane. The finite Point Spread Function (PSF) causes nearby sources to be blended. The focus-limited PSF (full width at half maximum of $\sim 1 -- 2$ pixel) and the large pixel size ($\sim$ 21\arcsec) of TESS imply that the resulting time-series photometry of a given target will often have contamination from nearby sources.

Blended light from nearby sources can decrease the depth, $\delta$, of a transit by
\begin{equation}
    \delta^\prime = \Big(1 - \dfrac{F_{\rm B}}{F_{\rm T} + F_{\rm B}}\Big) \delta = (1 - D) \delta = (1 - \dfrac{f}{1 + f}) \delta
    \label{equa:dilu}
\end{equation}
where $\delta^\prime$ is the diluted transit depth, $F_{\rm B}$ and $F_{\rm T}$ are the fluxes of the blended and target source, respectively. Here, $D$ is dilution, and $f\equiv F_{\rm B}/F_{\rm T}$ is the flux ratio of the blended and target objects. The SPOC pipeline corrects the PDC light curves for this dilution of the transits.

\begin{figure*}
    \centering
    \includegraphics[trim=1cm 8cm 0cm 7.5cm, clip, width=\textwidth]{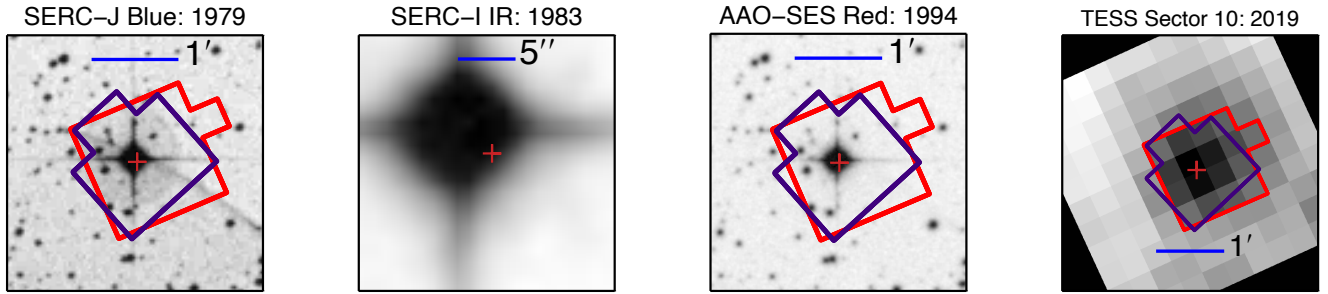}
    \caption{The archival and TESS images of HD~108236. The TESS image is from Sector 10 taken during 2019. Overplotted on the TESS image are the two apertures that are used to extract the light curves during Sector 10 (red) and 11 (purple).}
    \label{figr:imag}
\end{figure*}

The TESS image of HD~108236 from Sector 10 is shown in Figure~\ref{figr:imag} along with several archival images of the target including the Science and Engineering Research Council (SERC) J image taken in 1979, SERC-I image taken in 1983 and the Anglo-Australian Observatory Second Epoch Survey (AAO-SES) image taken in 1994. The apertures that are used to extract the TESS light curves are also shown for Sector 10 (red) and 11 (purple). Some of the relatively bright neighbors of HD~108236 are TIC 260647148, 260647113, 260647110, and 260647155 that are 77, 95, 108, and 122\arcsec away and have TESS magnitudes of 13.89, 13.73, 12.94, and 11.67, respectively. Due to the large aperture used to collect light from the bright target HD~108236, the total flux from blended sources is roughly $f=1.2\%$ of the photons coming from HD~108236. 

Detection of periodic transits in photometric time-series data can be due to any of the following:

\begin{itemize}
    \item An instrumental (systematic) effect,
    
    \item The primary (i.e., brightest) star being eclipsed by a companion star (i.e., eclipsing binary),
    
    \item A foreground or background star (i.e., gravitationally not associated with the target) \emph{aligned} with the target being eclipsed by a stellar companion or transited by a planet,
    
    \item The primary or one of the fainter (secondary) stars in a hierarchical multiple star system eclipsing each other or being transited by a planet,

    \item A nearby star (i.e., gravitationally not associated with the target) being eclipsed by a stellar companion or transited by a planet,
    
    \item A star being transited by a planet.
\end{itemize}

Therefore, we individually considered and ruled out the alternative hypotheses in order to ensure that the planetary classification for the origin of the detected transits was not a false positive.

The first false-positive hypothesis was that the transits could be due to an instrumental effect. The orbital periods of TOI~1233.03 and TOI~1233.04 were close to the multiples of the momentum dump period, which occurred every 3.125\,days for Sectors~10 and 11, according to the TESS Data Release Notes\footnote{https://archive.stsci.edu/tess/tess\_drn.html}. However, the detected transits did not fall near the momentum dumps. In addition, the transit shapes were inconsistent with that of the typical momentum dump artifact (i.e., sudden drop followed by a gradual rise). The difference images also did not show any evidence of scattered light in the vicinity of HD~108236 during the observations of interest. Furthermore, there were many individual transits detected, which made it extremely unlikely that they were produced by unrelated systematic events. This ruled out the instrumental origin of the detected transits.

The transit model fit performed by the SPOC pipeline on the TESS data indicated that the transit was not grazing and that the depth and shape of the transits were consistent with being of planetary nature. This was also confirmed later with our transit model as discussed in Section~\ref{sect:modl}. The SPOC data validation also showed that the apparent positions of the TCEs were all within 1 pixel of HD~108236. Nevertheless, the periodic dimming could be due to any of the sufficiently bright sources in the aperture, since transits or eclipses from nearby or physically associated companion stars could be blending into the aperture. In general, dynamical measurements such as Transit Timing Variations (TTVs) could break this degeneracy. However, the small number of transits and the limited baseline ($\sim$ 60 days) of the detection data did not yet allow TTVs to be used for vetting.

As a result, follow-up observations were needed to rule out the remaining false-positive hypotheses that the transits are on a target other than the brightest target (i.e., primary). In the remainder of this section, we summarize the data we collected to rule out these false positive hypotheses.

\subsection{Reconnaissance spectroscopy}
\label{sect:recospec}

Upon TESS detection, we obtained reconnaissance spectroscopy follow-up data on HD~108236 using the resources of the SG2 subgroup of TFOP at the Cerro Tololo Inter-American Observatory (CTIO) in Chile, including the Network of Robotic Echelle Spectrographs (NRES) of the Las Cumbres Observatory and the CTIO high-resolution spectrometer (CHIRON).

\subsubsection{LCO/NRES}
\label{sect:lcocnres}
The NRES \citep{Siverd+2016} instrument at Las Cumbres Observatory Global Telescope (LCOGT) \citep{Brown+2013} consists of four identical, high-precision spectrographs in the optical band (i.e., 390--860\,nm). We used LCO/NRES at the CTIO in Chile to collect two high-resolution spectra of HD~108236. Each one of these two observations consisted of three consecutive 20 minute stacked exposures. The raw data were then processed by the NRES data reduction pipeline, which included bias and dark corrections, optimal extraction of the one-dimensional spectrum, and the wavelength calibration with ThAr lamps. The resulting calibrated spectra were analysed using \texttt{SpecMatch}\footnote{\url{https://github.com/petigura/specmatch-syn}} \citep{Petigura2015, Petigura+2017}, by accounting for the Gaia parallax and using \texttt{Isoclassify} \citep{Huber+2017} to infer the physical parameters of the host star. Specifically, a 95\% confidence level upper bound of 2\,km$s^{-1}$ was placed on the sky-projected stellar rotation.

\subsubsection{SMARTS/CHIRON}

We observed HD~108236 with the CHIRON instrument \citep{Tokovinin+2013} mounted on the 1.5 meter Small and Moderate Aperture Research Telescope System (SMARTS) telescope at CTIO, Chile. We obtained 5 spectra using SMARTS/CHIRON on different nights. The exposure time was 100 seconds and each observation contained three exposures. We used the image slicer mode and obtained a spectral resolution of $R \sim 80,000$. No lithium absorption line was observed in the resulting spectra, indicating that the star is not young. Furthermore, no stellar activity was observed in the H$_\alpha$ line. The stellar characterization obtained based on the LCO/NRES and SMARTS/CHIRON data are shown in Table~\ref{tabl:starchar}.

\subsubsection{Ruling out aligned eclipses and transits}

The cross correlation function and the Least Squares Deconvolution (LSD) line profile inferred from the reconnaissance spectra rule out well-separated or even partially blended secondary set of lines, constraining any spatially blended companion with different systemic velocities to be fainter than 5\% of the primary at 3 $\sigma$ in the TESS band. This flux ratio is linked to the difference of the magnitudes of the blended source, $m_{\rm B}$, and the target source, $m_{\rm T}$, as
\begin{equation}
    m_{\rm B} - m_{\rm T} = -2.5 \log_{10} f,
    \label{equa:dmag}
\end{equation}
which implies that the SG2 data rule out spatially blended sources that have different systemic velocities and that are brighter than TESS magnitude 11.9. 

Furthermore, through transit geometry, the undiluted depth, $\delta \equiv (R_{\rm p}/R_\star)^2$, of a full (i.e., non-grazing) transit is linked to full and total transit durations. The total transit duration $T_{\rm tot}$ is the time interval during which at least some part of the transiting object is occluding the background star, whereas the full transit duration $T_{\rm full}$ is the time interval during which the transiting object is fully within the stellar disk. Therefore, modeling of the full and total transit durations based on the observed transits allows the estimation of dilution of a transit caused by its neighbors. We inferred the dilution consistent with the observed TESS transits using a methodology similar to that discussed in Section~\ref{sect:modl}. The marginal posterior of the dilution requires any blended source to be brighter than TESS magnitude 12.1 at 2 $\sigma$ to produce the observed TESS light curve. Therefore, combined with the constraint from the SG2 data, this rules out the hypothesis that the transits could be produced by a faint foreground or background binary. Furthermore, the fact that there are multiple TCEs on the same target implies that the alignment of unassociated background or foreground eclipses or transits are very unlikely \citep{Lissauer+2012}.

\begin{table}[]
    \centering
    \caption{SG2 and SG4 spectroscopic observations performed on HD~108236.}
    \begin{tabular}{m{3.5cm} | m{3.5cm}}
        \hline
        Telescope & SMARTS \\
        Instrument & CHIRON\\
        Spectral resolution [R] & 80,000\\
        Wavelength coverage& 4500 - 8900\,\AA\\
        SNR/resolution element & 44.2\\
        SNR wavelength & 5500\,\AA\\
        \hline
        Telescope & LCOGT \\
        Instrument & NRES \\
        Spectral resolution (R) & 48,000\\
        Wavelength coverage & 3800 - 8600\,\AA\\
        SNR/resolution element & 41.6\\
        SNR wavelength & 5500\,\AA\\
        \hline
        Telescope & Magellan II \\
        Instrument & PFS\\
        Spectral resolution [R] & 130000\\
        Wavelength coverage& 3800 - 6900\,\AA\\
        SNR/resolution element & 125 \\
        SNR wavelength & 5600\,\AA
    \end{tabular}
\end{table}

\subsection{Precise Doppler spectroscopy}

The reconnaissance spectroscopy data justified further follow-up of the target to obtain precise radial velocities using the SG4 resources of TFOP.

\subsubsection{Magellan II/PFS}

We used the Planet Finder Spectrograph (PFS) instrument \citep{Crane+2006, Crane+2008, Crane+2010} on the 6.5-meter Magellan II (Clay) telescope \citep{Johns+2012} at Las Campanas Observatory in Chile to obtain high-precision radial velocities of HD~108236 in July and August of 2019. PFS is an optical, high-resolution echelle spectrograph and uses an iodine absorption cell to measure precise radial velocities as described in \citet{Butler+1996}. We obtained a total of 12 radial velocity observations (with exposure times ranging from 15 to 20 minutes) and an iodine-free template observation of 30 minutes, yielding typical a precision of 0.64--1.5 m s$^{-1}$. Our PFS velocities are listed in Table~\ref{tabl:radvdata}. 

HD~108236 is also a target in the Magellan-TESS Survey (MTS; Teske et al., in prep), which measures precise masses of $\sim$30 planets with R$_{\rm p}<3$ R$_{\oplus}$ detected in the first year of TESS observations. Additional precise radial velocity observations made with PFS will be used to place constraints on the masses of the HD~108236 planets in the near future.

\subsubsection{Ruling out stellar companions}

Table~\ref{tabl:radvdata} summarizes the radial velocity measurements collected by the SG2 and SG4 subgroups of TFOP. The radial velocities obtained using NRES data are consistent with that from Gaia DR2 \citep{Brown+2018, Bailer-Jones+2018}, whereas radial velocities inferred from CHIRON observations have a systematic offset.

\begin{table}[]
    \centering
    \caption{Radial velocity data collected as part of reconnaissance (SG2) and precision (SG4) spectroscopy. }
    \begin{tabular}{c|c|c}
        \hline
        Time [BJD] & RV [km s$^{-1}$]	& 1$\sigma$ RV uncertainty  [km s$^{-1}$] \\
        \hline
        \hline
        NRES \\
        \hline
        2458647.567839 & 16.93 & 0.07 \\
        2458658.456917 & 16.82 & 0.11 \\
        \hline
        \hline
        CHIRON \\
        \hline
        2458666.59558 & 15.283 & 0.027 \\
        2458668.62232 & 15.385 & 0.027 \\
        2458698.51351 & 15.391 & 0.042 \\
        2458872.85177 & 15.416 & 0.036 \\
        2458876.83875 & 15.319 & 0.034 \\
        \hline
        \hline
        Time [JD] & DRV [m s$^{-1}$]	& 1$\sigma$ DRV uncertainty  [m s$^{-1}$] \\
        \hline
        \hline
        PFS \\
        \hline
        2458676.50493 &     5.31 &  0.68 \\
        2458679.53299 &    -1.25 &  0.84 \\
        2458680.53958 &    -0.21 &  0.80 \\
        2458682.51067 &     2.14 &  0.92 \\
        2458684.51457 &    -2.52 &  0.87 \\
        2458703.50490 &    -1.00 &  1.30 \\
        2458705.47891 &    -4.38 &  1.04 \\
        2458707.48948 &     2.00 &  1.08 \\
        2458709.49288 &    -1.73 &  1.01 \\
        2458713.49567 &    -1.85 &  1.25 \\
        2458716.47714 &     0.00 &  1.01 \\
        2458717.49043 &     4.66 &  1.50 \\
        \hline
    \end{tabular}
        \begin{minipage}{8cm}
            DRV: differential radial velocity
        \end{minipage}
    \label{tabl:radvdata}
\end{table}

\begin{figure*}
    \centering
    \includegraphics[trim=2cm 0.5cm 1cm 0cm, clip]{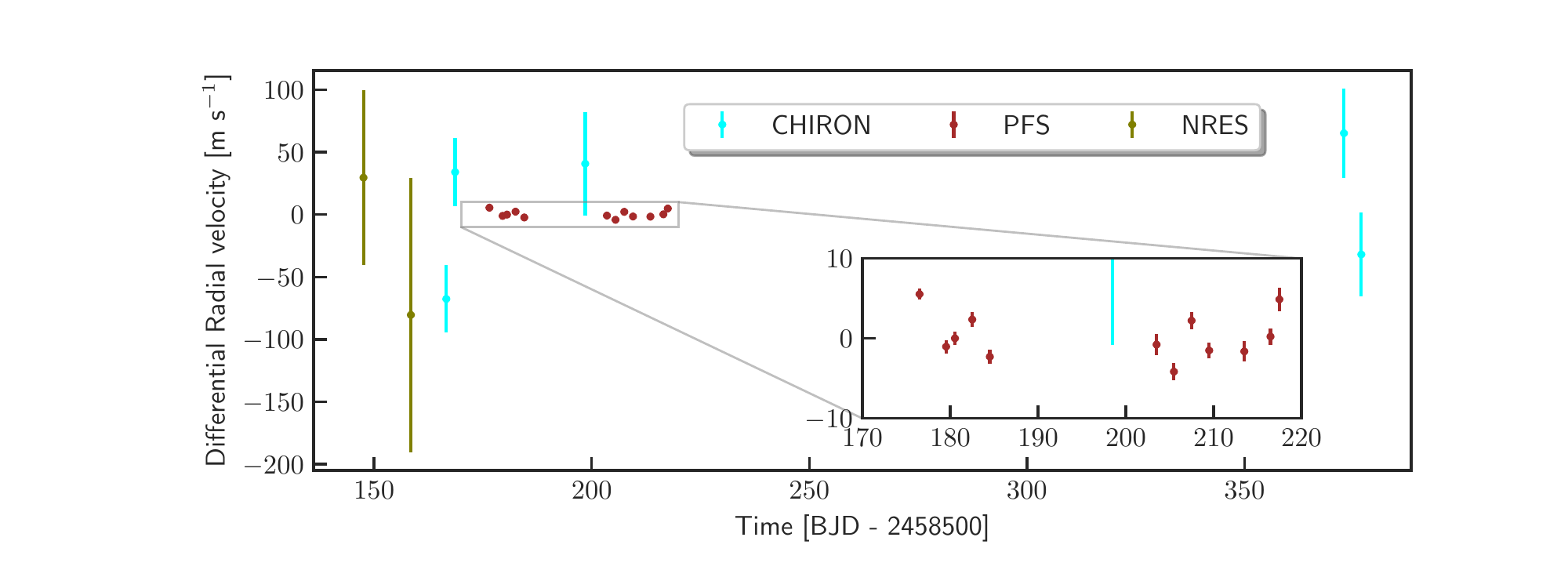}
    \caption{Differential radial velocities of HD~108236 measured as part of the SG2 and SG4 subgroups, modeled using a sinusoidal (i.e., non-eccentric) model. NRES, CHIRON, and PFS data are shown with the colors olive, cyan, and brown, respectively. The data from each instrument are shown after subtracting the weighted mean.}
    \label{figr:radv}
\end{figure*}

Figure~\ref{figr:radv} shows the radial velocity data from NRES, CHIRON and PFS after subtracting the mean within each data set. Among the three data sets, the PFS data have the smallest uncertainties ($\sim$ 1 m s$^{-1}$). However, they also display variations larger than the uncertainties. This is likely caused by the Doppler shifts due to planets validated in this work.

The root mean square (RMS) of the radial velocity data from NRES, CHIRON, and PFS are 55, 50, and 3 m s$^{-1}$, respectively. Using the RMS of the PFS radial velocity data, we can place a 3$\sigma$ upper limit of 1450 M$_\oplus$ on the mass of a companion on a circular orbit around HD~108236 with an orbital period less than 1000 days and an orbital inclination of 90 degrees. Furthermore, assuming circular orbits, the PFS data allow us to rule out stellar masses for the objects that have been observed by TESS to transit HD~108236. This is because the observed RMS of the PFS data is much smaller than the expected radial velocity semi-amplitude ($\sim$ 1 km s$^{-1}$) from a stellar object having a mass larger than $\sim$13.6 times the Jovian mass.

We note that we did not use the 12 precise radial velocity measurements from PFS to measure the masses of any of the four planets validated in this work. We leave this to a future work (Teske et al., in prep), where a larger set of precise radial velocity measurements from PFS will be used to accurately measure the masses of the validated planets.

The currently available radial velocity data cannot rule out stellar companions at arbitrary orbital periods, eccentricities and inclinations. Therefore, a remaining false positive hypothesis would be a hierarchical system containing planets transiting the primary or the secondary. However, the transiting planets would also have to be giants in this case, in order to compensate for the dilution from the companion star. If more than one such giant planets orbited the companion star, the system would be dynamically unstable. The multiplicity of the transiting objects in the system makes this false positive hypothesis unlikely. Furthermore, as has been shown in \citet{Latham+2011, Lissauer+2012, Guerrero+2020}, it is much less likely for a planet candidate to be a false positive in a multiplanetary system than in a system with a single planet. We therefore discarded this false positive hypothesis based on the observation of four independent TCEs.

\subsection{High-resolution speckle imaging}
\label{sect:highreso}
In order to rule out aligned foreground or background stars at close separations, high-resolution images are needed. To obtain high-resolution images in the presence of atmospheric scintillation, we used the speckle imaging technique by taking short exposures of the bright target to factor out the effect of atmospheric turbulence. For this purpose, we used the resources of the SG3 subgroup of TFOP and obtained high-resolution speckle images of HD~108236 with SOAR/HRCam and Gemini/Zorro.

\subsubsection{SOAR/HRCAM}

Diffraction-limited resolution was obtained via speckle interferometry by using the High-Resolution Camera (HRCam) \citep{Tokovinin+2010, Ziegler+2020} at the 4.1-meter SOAR telescope by processing short-exposure images taken with high magnification on UT 7 January 2020. The autocorrelation function and the resulting sensitivity curve are presented in the left panel of Figure~\ref{figr:spek}. A contrast of 5 magnitudes is achieved at a separation of 0\farcs2.

\subsubsection{Gemini/Zorro}

\begin{figure*}
    \centering
    \includegraphics[width=0.49\textwidth]{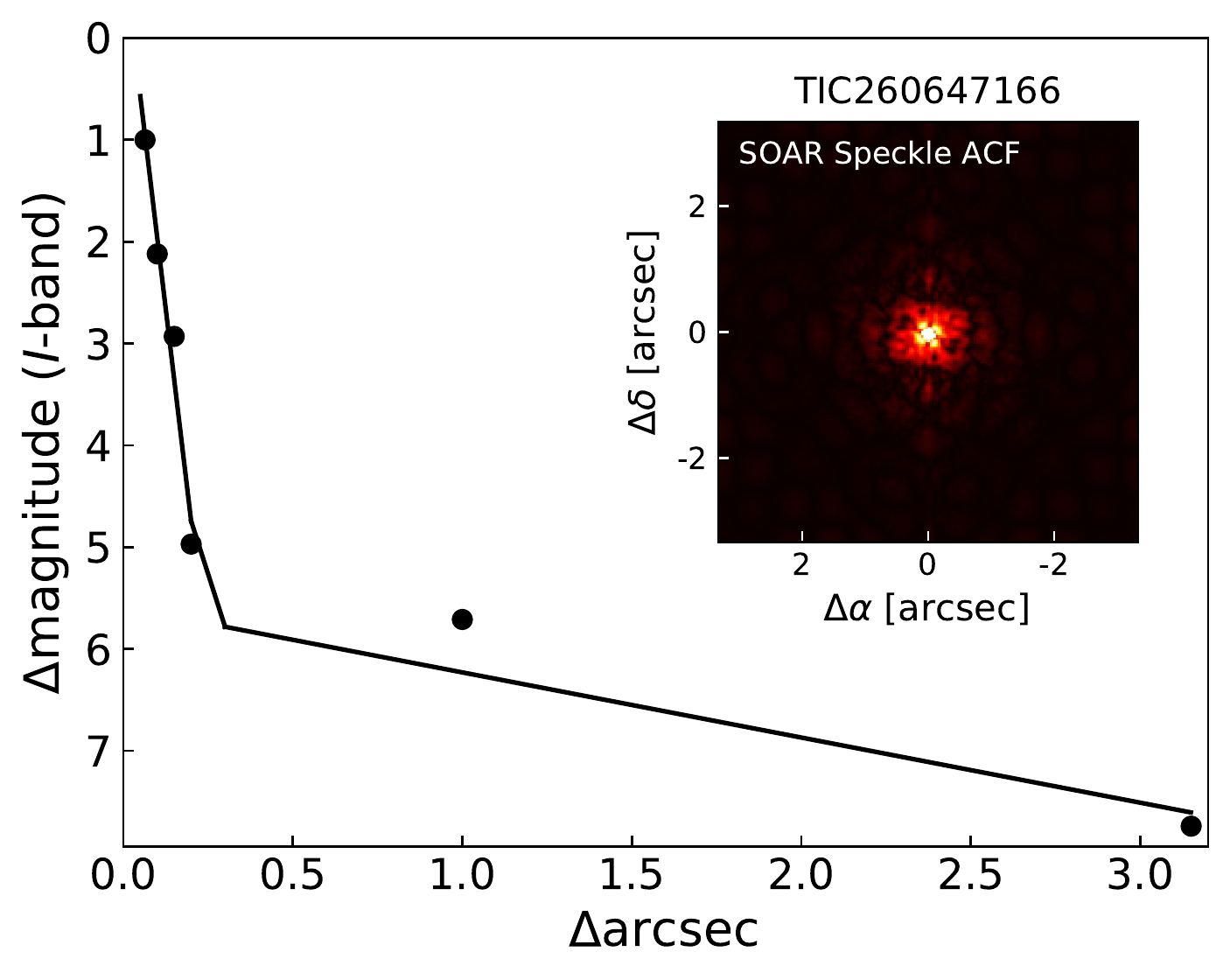}
    \includegraphics[width=0.49\textwidth]{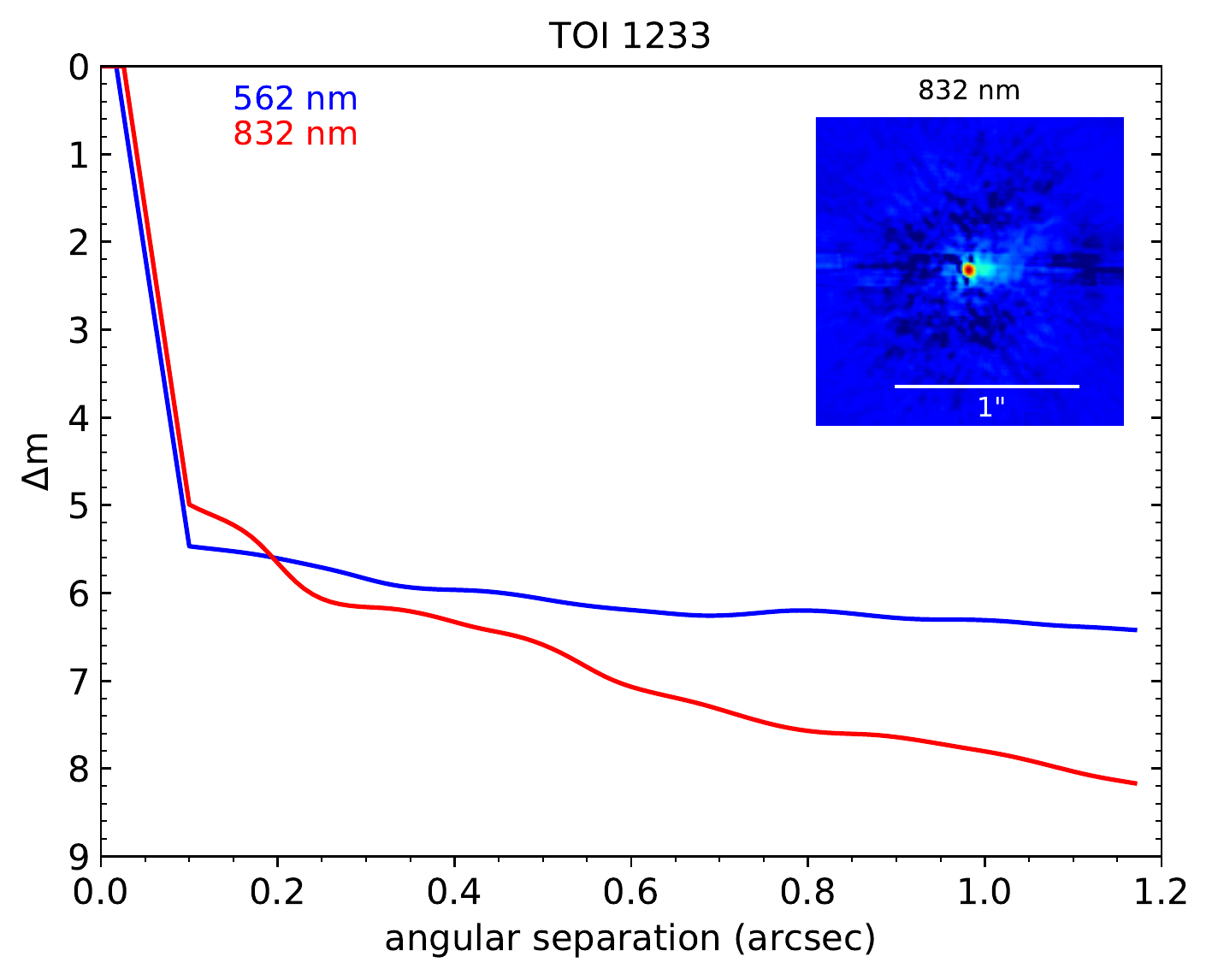}
    \caption{The 5-$\sigma$ sensitivity curve of speckle imaging by SOAR/HRCam (left) and Gemini/Zorro (right). The inset on the left shows the two-dimensional autocorrelation function, whereas the inset on the right is a reconstructed image of the field. The data rule out bright neighbors and companions to HD~108236, which would be fully spatially-blended in the TESS images.}
    \label{figr:spek}
\end{figure*}

We obtained speckle interferometric images of HD~108236 on UT 14 January 2020 and UT 12 March 2020 using the Zorro\footnote{https://www.gemini.edu/sciops/instruments/alopeke-zorro/} instrument on the 8-meter Gemini South telescope at the summit of Cerro Pachon in Chile. Zorro simultaneously observes in two bands, i.e., $832\pm40$\,nm and $562\pm54$\,nm, obtaining diffraction limited images with inner working angles of 0.017\arcsec and 0.026\arcsec, respectively. Both data sets consisted of 3 minutes of total integration time taken as sets of a thousand 0.06-second images. Each night's data were combined and subjected to Fourier analysis leading to the production of final data products including speckle reconstructed imagery. The right panel of Figure \ref{figr:spek} shows the 5-sigma contrast curves in both filters for data collected on UT 12 March 2020 and includes an inset showing the 832\,nm reconstructed image. The speckle imaging results in both observations agree, revealing HD~108236 to be a single star to contrast limits of 5.5 to 8 magnitudes within a sky-projected separation between 1.3 and 75 Astronomical Unit (AU).

These high-resolution images rule out wide stellar binaries that would not be spatially-resolved in ground-based, seeing-limited photometry with a PSF of $\sim$ 1\arcsec. In addition, using the Dartmouth isochrone model \citep{Dotter2008}, they imply that a bound companion to HD~108236 would have to be less massive than 0.10-0.15 M$_\odot$, depending on the age of the system.

\begin{table}[]
    \centering
    \caption{High-resolution imaging data collected on HD~108236.}
    \begin{tabular}{m{3.5cm} | m{3.5cm}}
        \hline
        Telescope & SOAR \\
        Instrument & HRCam\\
        Filter & $879 \pm 289$\,nm\\
        Image Type & Speckle\\
        Pixel Scale [as] & 0.01575\\
        Estimated PSF [as] & 0.06364\\
        \hline
        Telescope & Gemini \\
        Instrument & Zorro\\
        Filter & $832 \pm 40$\,nm, $562 \pm 54$\,nm \\
        Image Type & Speckle\\
        Pixel Scale [as] & 0.01\\
        Estimated PSF [as] & 0.02\\
        \hline
    \end{tabular}
\end{table}

\subsection{Seeing-limited (ground-based) transit photometry}

After ruling out binaries and chance alignments for the target, we then proceeded with ruling out the possibility that the transits detected by TESS could be on nearby stars. HD~108236 is the brightest source within a few arcminutes in its vicinity. Given the depth of the transits observed by TESS (\bdepthtrdilutedTESS{} ppt\footnote{We use ppt as a shorthand notation for parts per thousand.}, \cdepthtrdilutedTESS{} ppt, \ddepthtrdilutedTESS{} ppt, and \edepthtrdilutedTESS{} ppt), the transit depth would have to be deeper by a certain amount as given by Equations~\ref{equa:dilu} and \ref{equa:dmag} if the transit was not on HD~108236, but rather on a fainter nearby target. In order to rule out the hypothesis that \emph{any} of the transits could be on a nearby target, we collected seeing-limited (i.e., with a PSF full-width at half maximum of $\sim$ 1 as) photometric time-series data during a predicted transit for \emph{each} planet candidate (i.e., TOIs 1233.01, 1233.02, 1233.03 and 1233.04) using the resources of the SG1 subgroup of TFOP including the LCOGT and MEarth telescopes. Table~\ref{tabl:dataphot} lists these observations. As will be discussed in Section~\ref{sect:dete}, one of these observations (UT 17 March 2020) resulted in a tentative detection of a transit on target.

\subsubsection{LCOGT}

We used LCOGT \citep{Brown+2013} of 1-meter class telescopes to obtain ground-based transit light curves of all four planet candidates of HD~108236. We used the {\tt TESS Transit Finder}, which is a customized version of the \texttt{Tapir} software package \citep{Jensen2013}, to schedule our transit observations. Specifically, observations were taken from the CTIO and South African Astronomical Observatory (SAAO) LCOGT locations. Both telescopes are equipped with a $4096\times 4096$\,pixel Sinistro camera whose pixel scale is 0.389\arcsec, resulting in a $26'\times 26'$ field-of-view. We achieved a typical PSF FWHM of 2.3\arcsec, which is about 30 times smaller than the TESS PSF. Each image sequence was calibrated using the standard \texttt{BANZAI} pipeline \citep{McCully+2018} while the differential light curves of HD~108236 and its neighbouring sources were derived using the \texttt{AstroImageJ} software package \citep{Collins+2017}.

Table~\ref{tabl:dataphot} summarizes our eight successful transit observations from LCOGT taken between UT 11 January 2020 and UT 17 March 2020. Explicitly, we collected data during two, three, two, and one transits of TOIs 1233.01, 1233.02, 1233.03, and 1233.04, respectively. All light curves were obtained with either 20 or 60-second exposures in either the $y$ or $z_{\rm s}$ bands to optimize photometric precision. Photometric apertures were selected by the individual SG1 observer based on the FWHM of the target's PSF in order to maximize the photometric precision. In each light curve we tested all bright neighbouring sources within $2.5'$ of HD~108236. Then we either tentatively detected the expected transit event on the target (i.e., on UT 17 March 2020 with LCOGT-CTIO) or were able to rule out transit-like events on all nearby targets down to the faintest neighbor magnitude contrasts reported in Table~\ref{tabl:dataphot} (i.e., during all other observations). For each planet candidate, all known Gaia DR2 stars within 2.5 arcminutes of HD~108236 that are bright enough to cause the TESS detection were ruled out as possible sources of the TESS detections.

\subsubsection{MEarth-South}

\begin{figure*}
    \centering
    \includegraphics[trim=0cm 3cm 1.5cm 1cm, clip, width=0.49\textwidth]{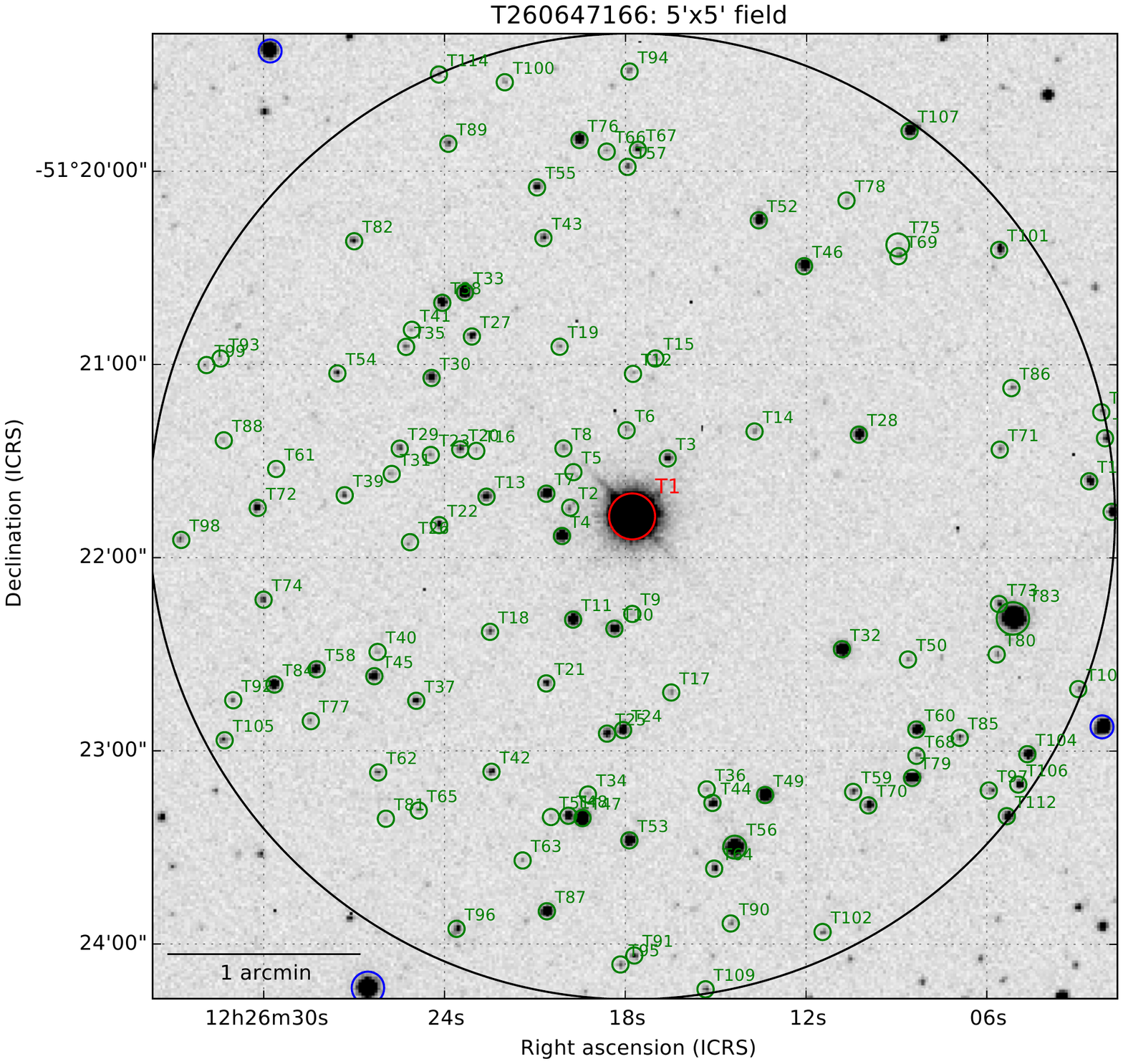} 
    \includegraphics[trim=0cm 3cm 1.5cm 0cm, clip, width=0.49\textwidth]{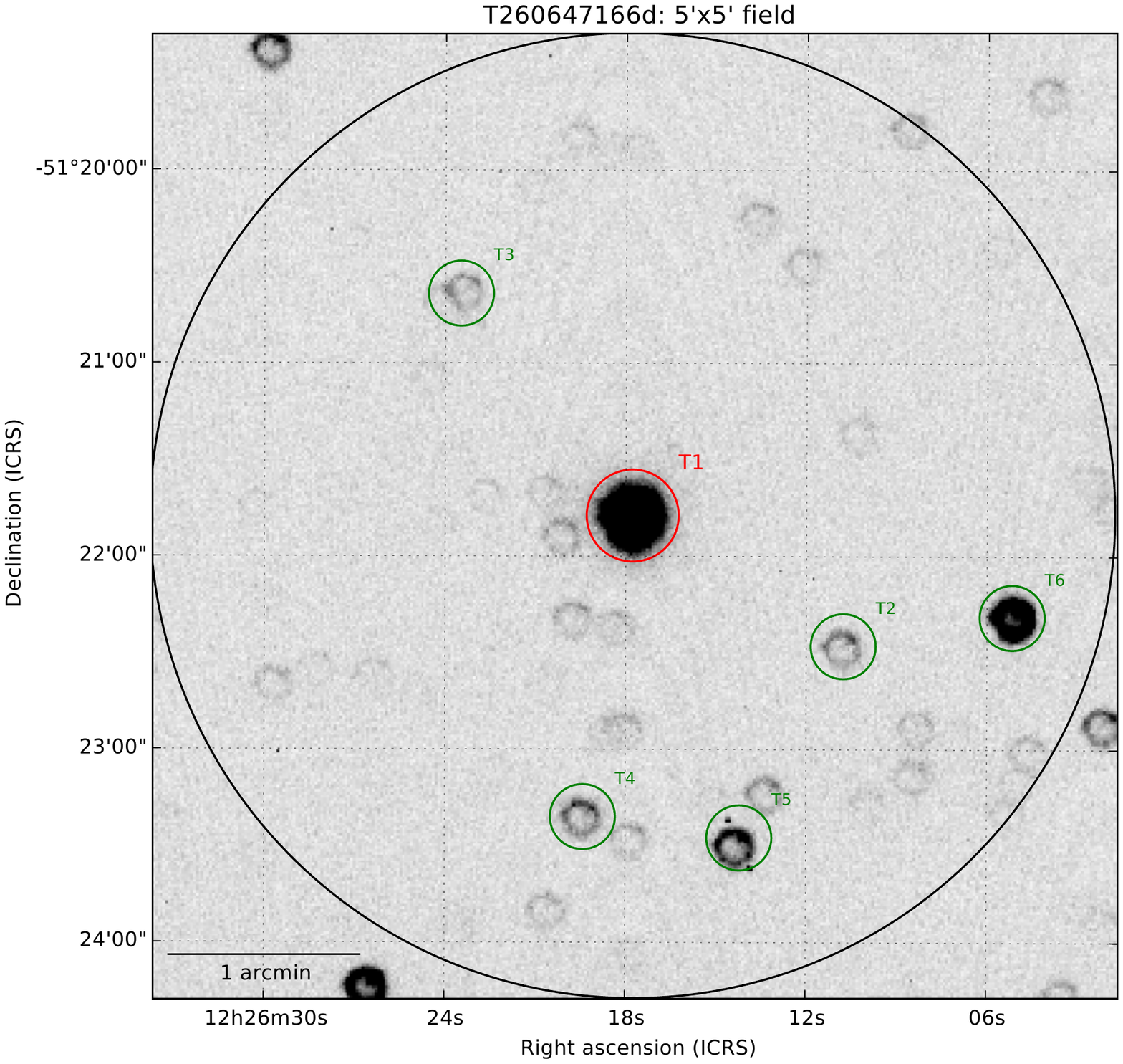}
    \caption{A photometric image of the field in the vicinity of HD~108236 as observed by MEarth-South on UT 17 March 2020. The left panel shows the image in focus as collected by one of the MEarth-South telescopes, where HD~108236 is saturated due to its brightness. The right panel shows the defocused image as observed by the other six MEarth-South telescopes. In these images the PSF is broader, unsaturating HD~108236 and allowing precision photometry on the target.}
    \label{figr:imagmear}
\end{figure*}

MEarth-South employs an array of eight f/9 40-cm Ritchey–Chrétien telescopes on German equatorial mounts \citep{Irwin+2015}. During the data acquisition for this work, only seven of the telescopes were operational. Data were obtained on three nights: UT 3 March 2020 (egress of TOI~1233.01), UT 11 March 2020 (full transits of TOI~1233.02 and TOI~1233.03) and UT 17 March 2020 (full transit of TOI~1233.01). Figure~\ref{figr:imagmear} shows the in-focus and defocused fields of the MEarth-South observation on UT 17 March 2020.

All observations were conducted using the same observational strategy. Exposure times were 35 seconds with six telescopes defocused to half flux diameter of 12 pixels to provide photometry of the target star, and one telescope observing in-focus with the target star saturated to provide photometry of any nearby or faint contaminating sources not resolved by the defocused time series. Observations were gathered continuously starting when the target rose above 3 air masses (first observation) or evening twilight (other observations) until morning twilight. Telescope 7 used in the defocused set had a stuck shutter resulting in smearing of the images during readout, but this did not appear to affect the light curves. The defocused observations were performed with a pixel scale of 0.84\arcsec. A photometric aperture with a radius of 17 pixels was used to extract the photometric time-series. Data were reduced following standard procedures for MEarth photometry \citep{Irwin+2007}.

\subsubsection{Ruling out nearby eclipses and transits}

During the predicted transit of each planet candidate (i.e., TOI~1233.01, TOI~1233.02, TOI~1233.03, and TOI~1233.04), light curves of all nearby stars were extracted and checked for any transits with a depth that could cause the relevant transits in the TESS light curves. No such transit was observed for any of the planet candidates. These data ruled out the hypotheses that any of the transits detected by TESS could be off-target by ensuring that no nearby star transited at the predicted transit time. 

Upon collecting the above time-series and ruling out transits on nearby targets, we finally concluded that the planetary nature of the transiting objects were validated. Thus, in the remainder of this paper, we will refer to these transiting planets as HD~108236\,b, HD~108236\,c, HD~108236\,d, and HD~108236\,e, (or simply as planet b, c, d, and e) ordered with respect to increasing distance from the host star, HD~108236. Note that these four planets correspond to TOIs 1233.04, 1233.03, 1233.01, and 1233.02, respectively.

\subsubsection{Ground-based detection of a transit}
\label{sect:dete}
\begin{figure}
    \centering
    \includegraphics[width=0.5\textwidth, trim=3cm 1cm 0cm 0cm, clip]{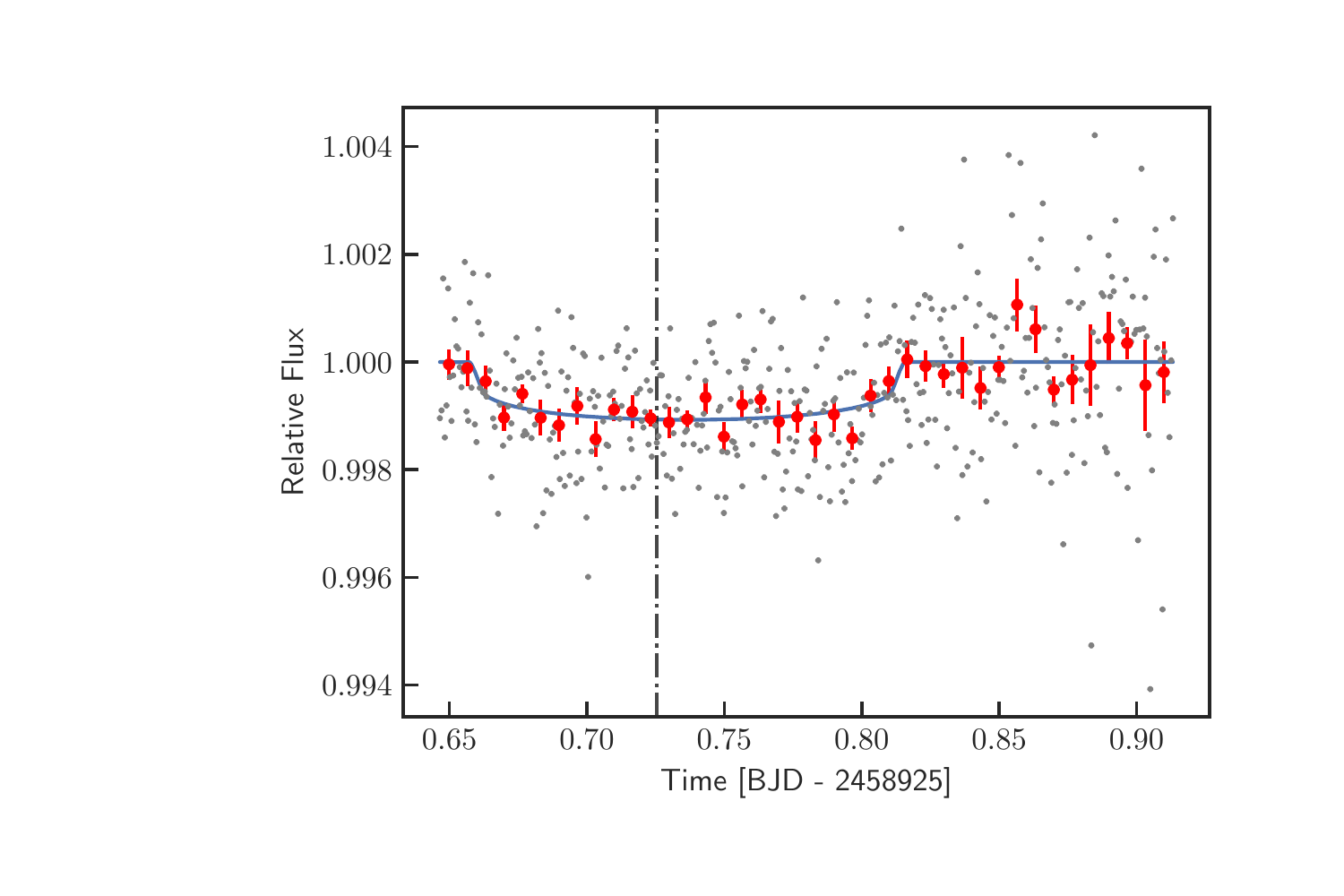}
    \caption{Detrended follow-up light curve of HD~108236 during the transit of planet d as measured by LCOGT-CTIO, where the transit was tentatively confirmed to be on-target. The transit arrived 14 minutes late, which is expected given the ephemeris uncertainty of $\sim$ 1 hour. The vertical line shows the mid-transit time of the transit that was expected based on the linear ephemeris inferred from the TESS data. The gray and red points denote the raw and binned data and the blue line is the posterior median transit model.}
    \label{figr:lcurtfop}
\end{figure}

A transit of planet d was tentatively detected on UT 17 March 2020 at a 1-meter LCOGT-CTIO telescope. The photometric time-series data had a relatively short pre-transit baseline. Therefore, we excluded these observations from the global orbital model in Section~\ref{sect:modl}, in order to avoid biasing the fit. However, we fitted the LCOGT-CTIO data separately and inferred a transit duration of $3.8\pm0.2$ hours and a transit depth of $1.1\pm0.2$ ppt, which are consistent with those inferred from the TESS data. The inferred mid-transit time was $2458571.3365\pm0.0035$ BJD, indicating a transit arrival 14 minutes late compared to the linear ephemeris model based on the TESS data. The associated light curve is shown in Figure~\ref{figr:lcurtfop}.

\begin{table*}[]
    \centering
    \caption{Ground-based photometric time-series observations made on HD~108236 during the predicted transits based on the TESS TCEs.}
    \medskip
    \raggedright
    
    \begin{tabular}{c|c|c|c|c|c|c|c|c|c|c|c}

    Date & Telescope & Camera & Filter & Pixel & PSF & AR & Transit & FN  & Duration  & Obs \\
    
    [UT] & & & & [as] & [as] & [Pixel] & & [Mag] & [minutes] & \\
    \hline
    \underline{\text{TOI~1233.01}} \\
  
    2020-03-02 & LCOGT-SAAO-1m & Sinistro & zs & 0.39 & 2.0 & 20 & Full & 8.1 & 341 & 376 \\
    \hline
    2020-03-03 & MEarth-South & Apogee & RG715 & 0.84 & 2.1 & 8.5 & Egress & 9.9 & 587 & 577 \\
    \hline
    2020-03-03 & MEarth-Southx6 & Apogee & RG715 & 0.84 & 8.0 & 17 & Egress & 5.5 & 588 & 3621 \\
    \hline
    2020-03-17 & LCOGT-CTIO-1m & Sinistro & zs & 0.39 & 2.5 & 15 & Full & n/c & 384 & 434 \\
    \hline
    2020-03-17 & MEarth-South & Apogee & RG715 & 0.84 & 2.1 & 8.5 & Full & 9.9 & 620 & 608 \\
    \hline
    2020-03-17 & MEarth-Southx6 & Apogee & RG715 & 0.84 & 8.1 & 17 & Full & 5.5 & 620 & 3819 \\
    \hline
    \hline
    \underline{\text{TOI~1233.02}}\\
    
    2020-01-11 & LCOGT-CTIO-1m & Sinistro & y & 0.39 & 1.8 & 10 & Ingress & 8.0 & 223 & 148 \\
    \hline
    2020-01-31 & LCOGT-SAAO-1m & Sinistro & y & 0.39 & 2.6 & 15 & Egress & 8.3 & 309 & 174 \\
    \hline
    2020-03-11 & MEarth-Southx6 & Apogee & RG715 & 0.84 & 7.9 & 17 & Full & 5.5 & 610 & 3759 \\
    \hline
    2020-03-11 & MEarth-South & Apogee & RG715 & 0.84 & 1.9 & 8.5 & Full & 11 & 609 & 584 \\
    \hline
    2020-03-11 & LCOGT-CTIO-1m & Sinistro & zs & 0.39 & 2.0 & 11 & Full & 7.7 & 455 & 507 \\
    \hline
    \hline
    
    \underline{\text{TOI~1233.03}}\\

    2020-02-02 & LCOGT-SAAO-1m & Sinistro & zs & 0.39 & 3.1 & 10 & Full & 8.6 & 296 & 192 \\
    \hline
    2020-03-11 & LCOGT-CTIO-1m & Sinistro & zs & 0.39 & 1.8 & 15 & Full & n/c & 452 & 507 \\
    \hline
    \hline
    \underline{\text{TOI~1233.04}}\\
    
    2020-01-11 & LCOGT-SAAO-1m & Sinistro & zs & 0.39 & 3.0 & 6 & Full & 9.2 & 205 & 143 \\
    
    \hline
    \end{tabular}
    \begin{minipage}{16cm} 
    FN stands for the faintest neighbor and the column values indicate the magnitude difference of the faintest neighbor checked for an NEB. In this column, (n/c) indicates "not checked" since transit-like events on nearby targets in the field at the same ephemeris were ruled out previously.
    \end{minipage}
    
    \label{tabl:dataphot}
\end{table*}

\subsection{Archival ground-based photometry}

HD~108236 has also been observed by the Wide Angle Search for Planets South (WASP-South) survey \citep{Pollacco+2006} in SAAO, South Africa. WASP-South, an array of 8 wide-field cameras, was the Southern station of the WASP transit-search project \citep{Pollacco+2006}. It observed the field of HD~108236 in 2011 and 2012, when equipped with 200-mm, f/1.8 lenses, and then again in 2013 and 2014, equipped with 85-mm, f/1.2 lenses. It observed on each clear night, over a span of 140 nights in each year, with a typical 10-minute cadence, and accumulated about 58,000 photometric measurements on HD~108236. We searched the data for any rotational modulation using the methods from \citet{Maxted+2011}. We found no significant periodicity between 1 and 80 days, with a 95\% confidence upper limit on the amplitude of 1 mmag. 
We did not detect any transits in the WASP data, consistent with the expected small transit depths of \bdepthtrundilutedTESS{}, \cdepthtrundilutedTESS{}, \ddepthtrundilutedTESS{}, and \edepthtrundilutedTESS{} ppt. Planet e had the deepest expected transit, however its relatively long period likely precluded any detection. The shallow transits of the inner planets also made them undetectable. To determine which region of the parameter space of transiting planets can be ruled out with the WASP data set, we performed injection-recovery tests using \texttt{allesfitter}, which will be introduced in Section~\ref{sect:modl}. We injected planets over a grid of periods of 10.1, 15.1, ..., 140.1\,days and radii of 8, 8.5, ..., 22\,\rearth{}. For each planet, we tried to recover the injected signal using Transit Least Squares \citep[\texttt{TLS},][]{HippkeHeller2019}. We find that $\sim$50\% of transiting planets with radii 1.3--2\,\rjup{} and periods less than 100 days could have been found in the WASP data. The recovery rate drops to $\sim$20\% for planets with radii $\sim$1\,\rjup{} and periods less than 100 days. In contrast, planets much smaller than Jupiter or those on periods longer than 100 days would remain undetected in the WASP data.

\subsection{Transit model}
\label{sect:modl}

Following the vetting of the planet candidates, we modeled the TESS PDC light curve to infer the physical properties of the orbiting planets. In order to model the photometric time-series data, we used \texttt{allesfitter} \citep{GuentherDaylan2019a, GuentherDaylan2019b}. The parameters $\theta$ of our forward model $M$ are presented in Table~\ref{tabl:paraalle}. We assumed a transit model with a linear ephemeris. We assumed a generic, eccentric orbit. For limb darkening, we used a transformed basis $q_1$ and $q_2$ of the linear $u_1$ and quadratic $u_2$ coefficients as \citep{Kipping2013} 
\begin{align}
    q_1 = (u_1 + u_2)^2, \\
    q_2 = 0.5 \dfrac{u_1}{u_1 + u_2}.
\end{align}

We modeled this red noise along with any other stellar variability in the data using a Gaussian Process (GP) with a Mat{\'e}rn 3/2 kernel as implemented by \texttt{celerite} \citep{Foreman-Mackey2017}.

\begin{table*}[]
    \centering
    \caption{Parameters of the transit forward-model.}
    \begin{tabular}{|c|c|c|}
        \hline
        Parameter & Explanation & Prior\\
        \hline
        \hline
        $q_{1; \mathrm{TESS}}$            & First limb darkening parameter 1  & uniform \\ 
        $q_{2; \mathrm{TESS}}$            & Second limb darkening parameter 2 & uniform \\ 
        $\log{\sigma_\mathrm{TESS}}$      & Logarithm of the scaling factor for relative flux uncertainties & uniform \\ 
        $\log{\sigma_{\rm GP;\mathrm{TESS}}}$ & Amplitude of the Gaussian process Mat{\'e}rn 3/2 kernel & uniform \\ 
        $\log{\rho_{\rm GP;\mathrm{TESS}}}$   & Time scale of the Gaussian process Mat{\'e}rn 3/2 kernel & uniform \\ 
        \hline
        $D_\mathrm{0; TESS}$              & Dilution of the transit depth due to blended light from neighbors & fixed \\ 
        \hline
        $R_{\rm n} / R_\star$                   & Ratio of planet $n$, $R_{\rm n}$, to the radius of the host star, $R_\star$ & uniform \\
        $(R_\star + R_{\rm n}) / a_{\rm n}$           & Sum of the stellar radius $R_\star$ and planetary radius $R_{\rm n}$ & uniform \\ 
        $\cos{i_{\rm n}}$                       & cosine of the orbital inclination, $i$ & uniform \\ 
        $T_{0;\rm n}$                         & Mid-transit time about which the linear ephemeris model pivots, i.e., epoc, in $\mathrm{BJD}$ & uniform \\ 
        $P_{\rm n}$                             & Orbital period of planet $n$ in days & uniform \\ 
        $\sqrt{e_{\rm n}}$ $\cos{\omega_{\rm d}}$       & Square root of the eccentricity times the cosine of the argument of periastron & uniform \\ 
        $\sqrt{e_{\rm n}}$ $\sin{\omega_{\rm d}}$       & Square root of the eccentricity times the sine of the argument of periastron & uniform \\ 
        \hline
    \end{tabular}
    \label{tabl:paraalle}
\end{table*}

When modeling the TESS data, we use the PDC light curve data product from the SPOC pipeline. We provide the posterior in Table~\ref{tabl:parafittnuis} for nuisance parameters, Table~\ref{tabl:parafittfrst} for the parameters of planets b and c, and Table~\ref{tabl:parafittseco} for the parameters of planets d and e. We then provide the derived posterior in Table~\ref{tabl:parainfefrst} for planets b and c and Table~\ref{tabl:parainfeseco} for planets d and e. Although our nominal results come from \texttt{allesfitter}, we have also repeated the analysis using \texttt{EXOFASTv2} \citep{Eastman+2019} as a cross check in order to confirm consistency. \texttt{EXOFASTv2} has a dynamical prior that avoids orbit crossings and ensures dynamical stability of the analyzed system. A notable result from this analysis were additional constraints on the eccentricities of the planets enabled by the Hill stability prior. The inferred eccentricities were smaller than 0.287, 0.197, 0.164, and 0.149 at a confidence level of 2$\sigma$ for planets b, c, d, and e, respectively.

\begin{figure*}[!htbp]
    \centering
    \includegraphics[trim=0.5cm 1cm 1cm 1cm, clip, width=\textwidth]{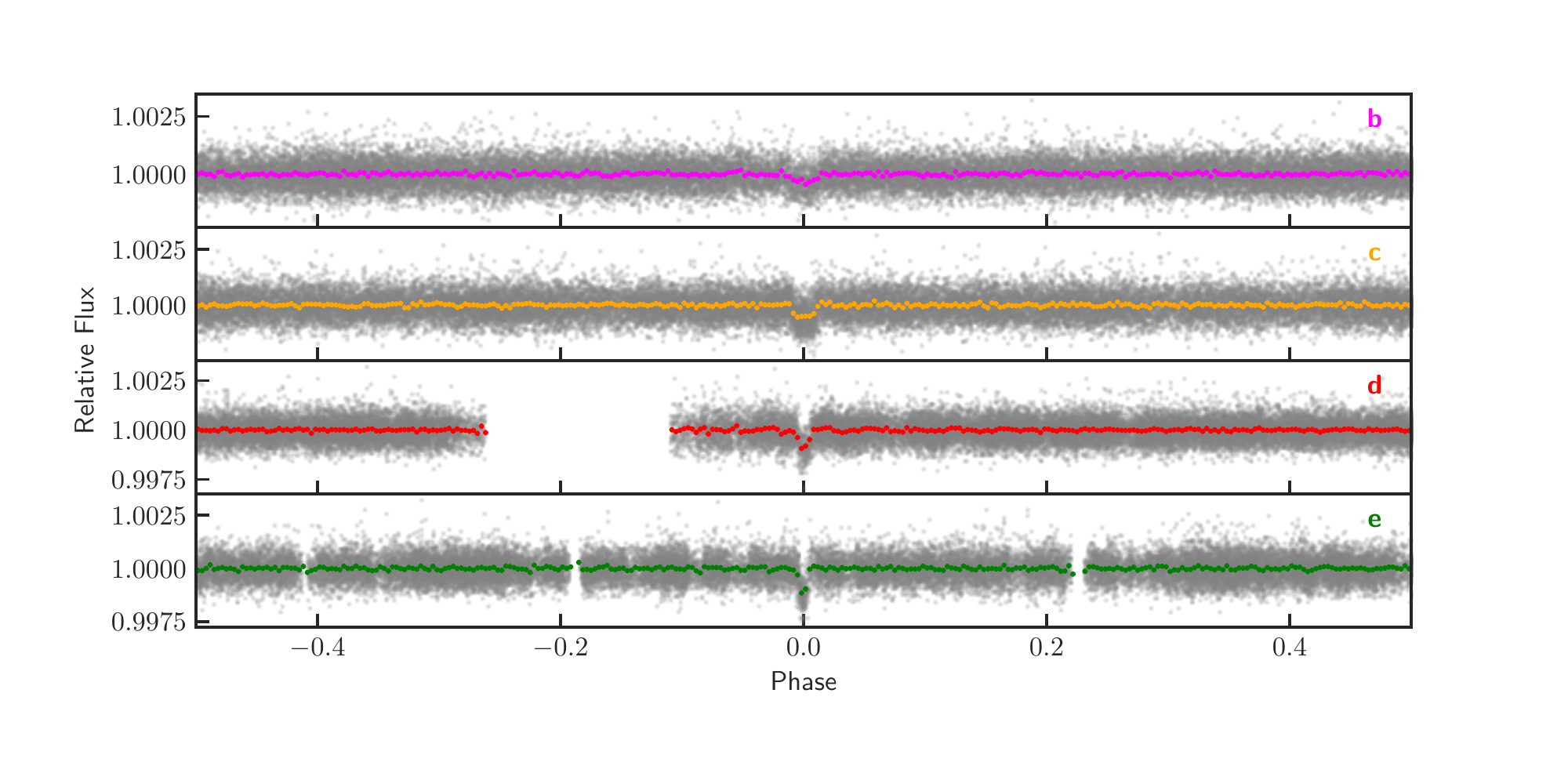}
    \caption{Detrended PDC light curve folded at the posterior median period of each planet after masking out the transits of other planets. Close-in views of the transits are also given in Figure~\ref{figr:pcur}.}
    \label{figr:pcur_post}
\end{figure*}

\begin{figure*}[!htbp]
    \centering
    \includegraphics[trim=2cm 1cm 1cm 1cm, clip, width=0.49\textwidth]{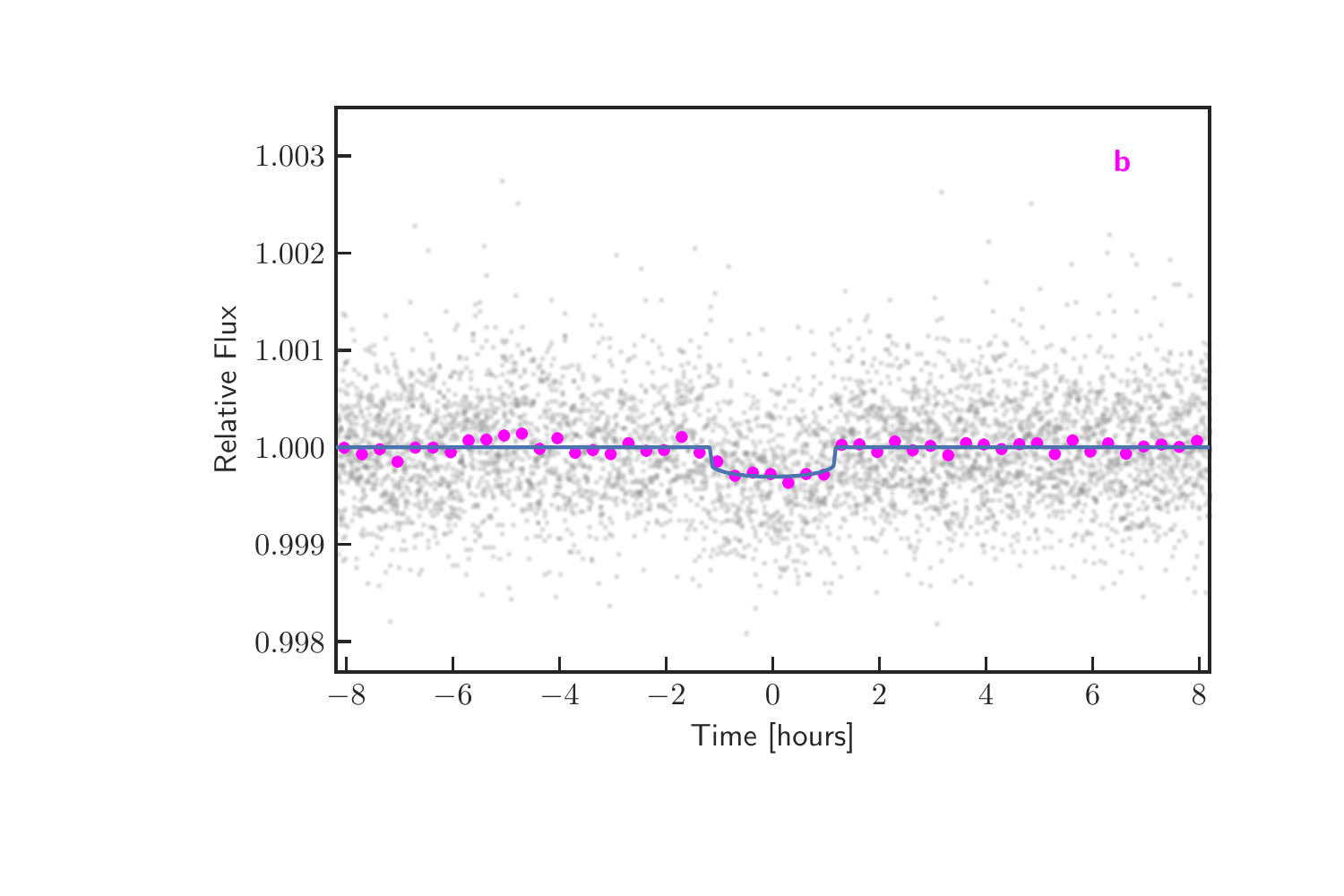}
    \includegraphics[trim=2cm 1cm 1cm 1cm, clip, width=0.49\textwidth]{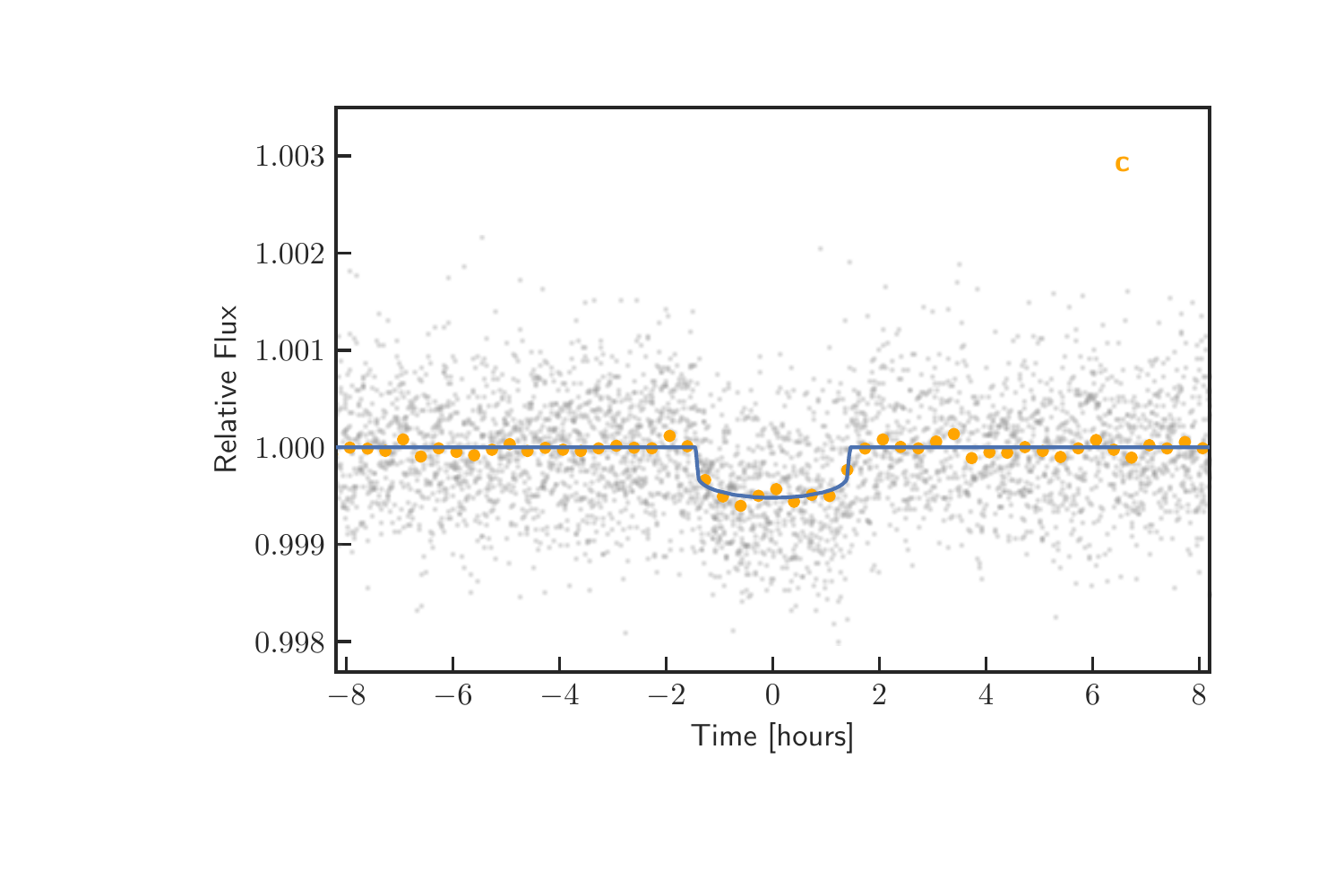} \\
    \includegraphics[trim=2cm 1cm 1cm 1cm, clip, width=0.49\textwidth]{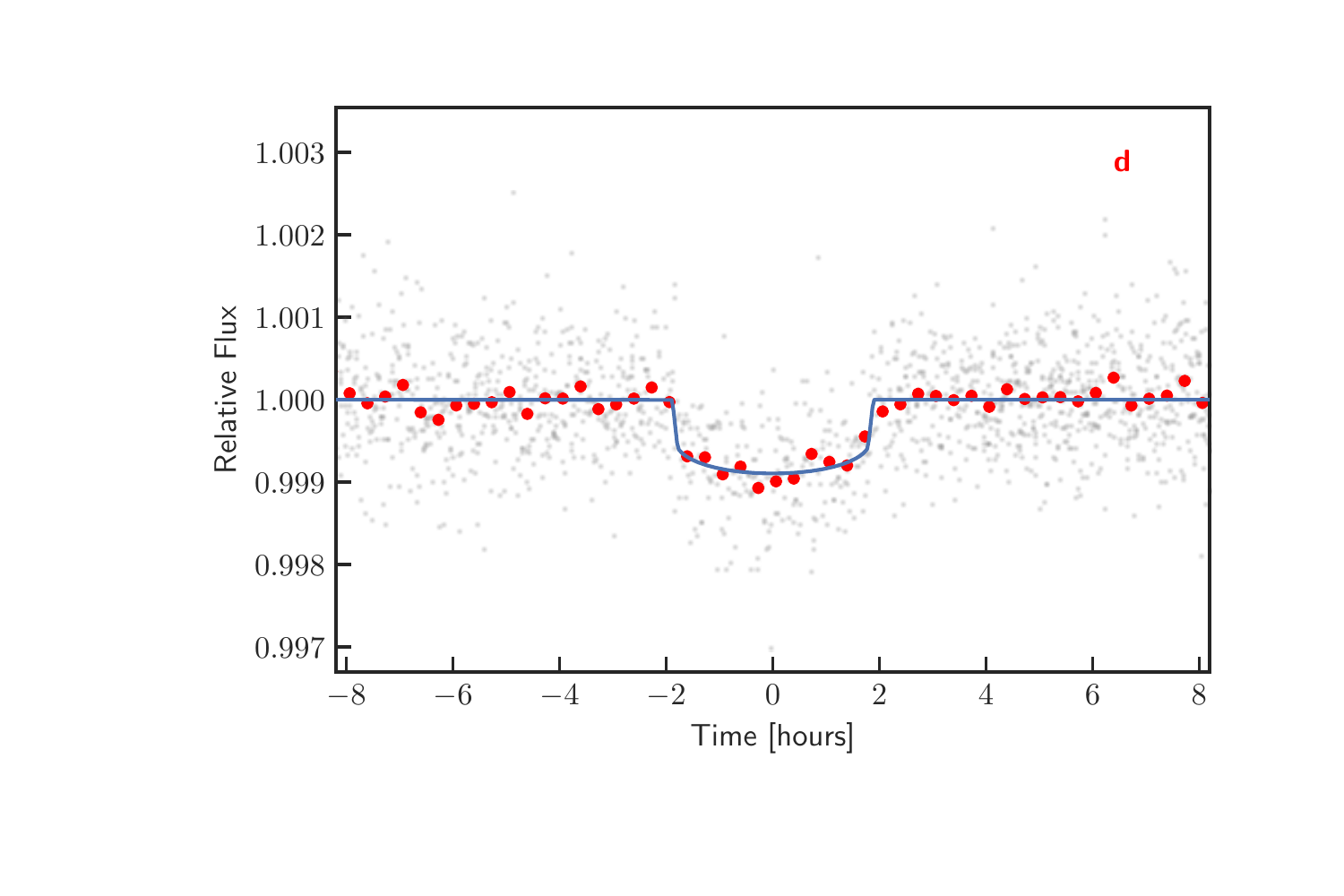}
    \includegraphics[trim=2cm 1cm 1cm 1cm, clip, width=0.49\textwidth]{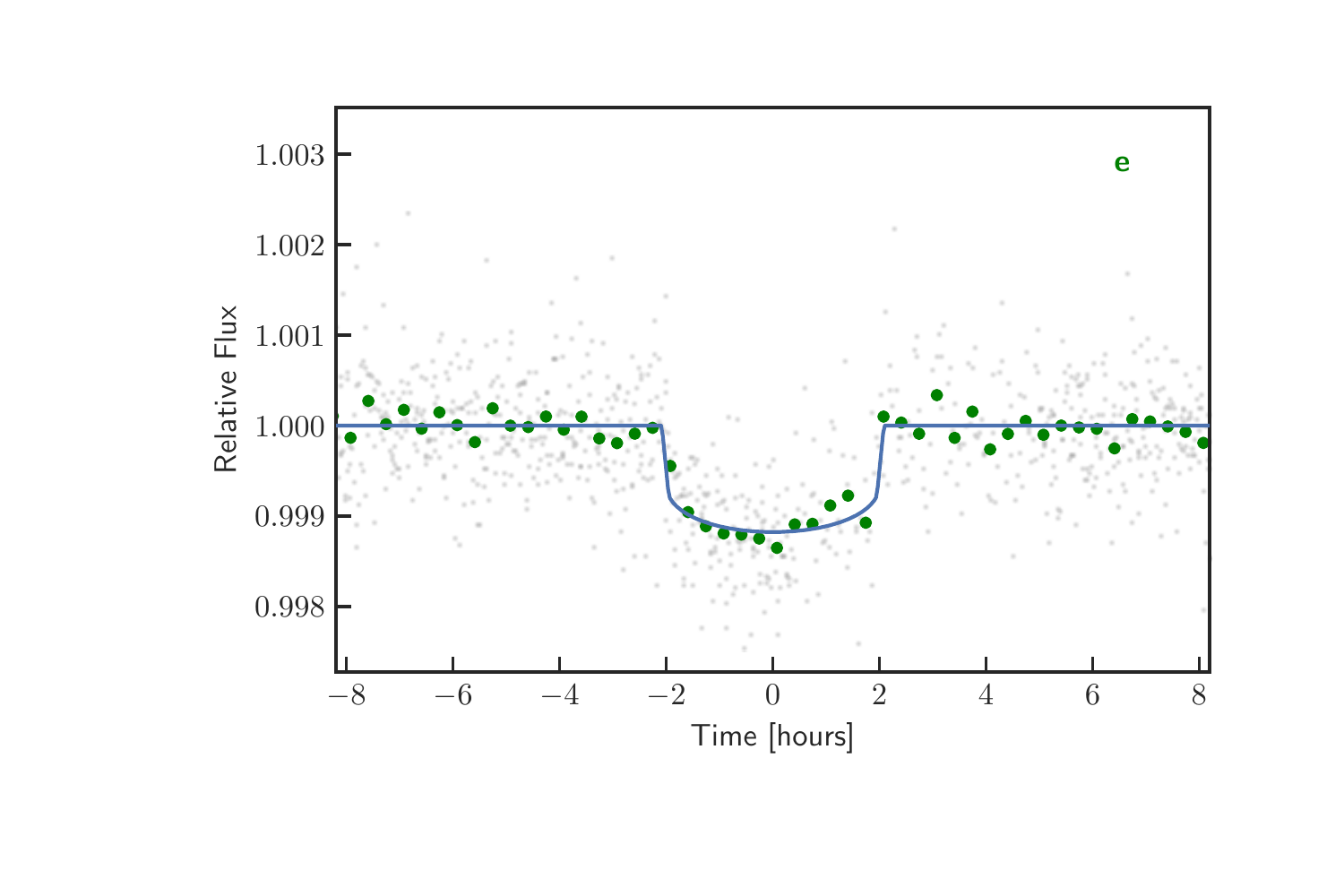}
    \caption{Phase curves of the four discovered planets. Blue lines indicate the posterior median of the transit model fitted to the data.}
    \label{figr:pcur}
\end{figure*}

We show in Figure~\ref{figr:pcur_post} the light curve of each planet folded onto its orbital period and centered at the phase of the primary transit, after masking out the transits of the other planets. Because the orbital period of planet d is close to the orbital period of TESS around the Earth ($\sim$ 13.7 days), a large gap is formed in its phase curve. Figure~\ref{figr:pcur} then shows the individual phase curves, along with the posterior-median transit model shown with the blue lines.
 
\section{The HD~108236 system}
\label{sect:plan}

\begin{figure*}
    \centering
    \includegraphics[trim=2cm 2cm 1cm 2cm, clip]{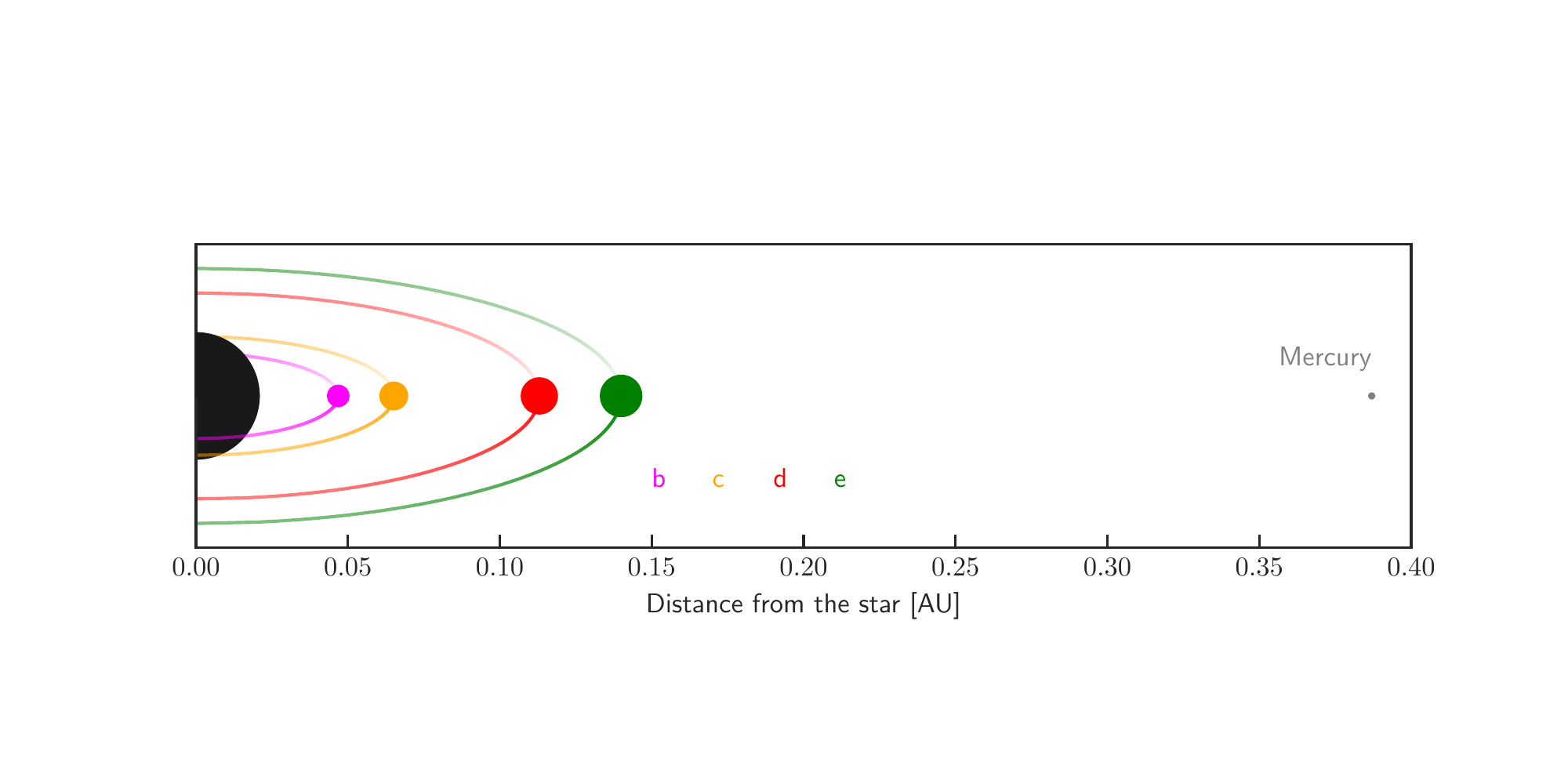}
    \caption{Inclined view of the HD~108236 system. The horizontal axis denotes the distance from the host star, HD~108236, which is shown on the left with a black circle. The four planets HD~108236\,b, HD~108236\,c, HD~108236\,d and HD~108236\,e are shown with magenta, orange, red and green, respectively. Shown on the far right with gray is Mercury as it would look if it orbited HD~108236 at its current orbital period. The radii of the planets and the star are scaled up by a factor of 50 and 5, respectively. The elliptical appearance of the orbits are due to the viewing angle and do not make any implication about the orbital eccentricities.}
    \label{figr:orbt}
\end{figure*}

In this section, we review the main properties of the planets discovered to be transiting HD~108236. The HD~108236 system is depicted in Figure~\ref{figr:orbt}. The transiting planets b, c, d, and e orbit the host star on orbits with semi-major axes of \baAU{} AU, \caAU{}, \daAU{} AU, and \eaAU{} AU, respectively. Compared to our Solar System, the discovered planets orbit rather closer to their host star, HD~108236, forming a closely-packed, compact multiplanetary system. 

HD~108236\,b is a hot super-Earth with a radius of \bRcompanionRearth{} $R_\oplus$. Being the innermost discovered planet of the system, it has a period of \bperiod{}\,days, making it the hottest known planet in the system with an estimated equilibrium temperature of \bTeq{}\,K. The other three known planets in the system are HD~108236\,c, HD~108236\,d, and HD~108236\,e. These are sub-Neptunes with radii \cRcompanionRearth{} $R_\oplus$, \dRcompanionRearth{} $R_\oplus$, and \eRcompanionRearth{} $R_\oplus$ and periods \cperiod{} days, \dperiod{} days, and \eperiod{} days, respectively. Their equilibrium temperatures are \cTeq{} K, \dTeq{} K, and \eTeq{} K, respectively, under the assumption of an albedo of 0.3.

\begin{figure}[!htbp]
    \centering
    \includegraphics[trim=0.5cm 0.5cm 0.5cm 0.5cm, clip, width=0.49\textwidth]{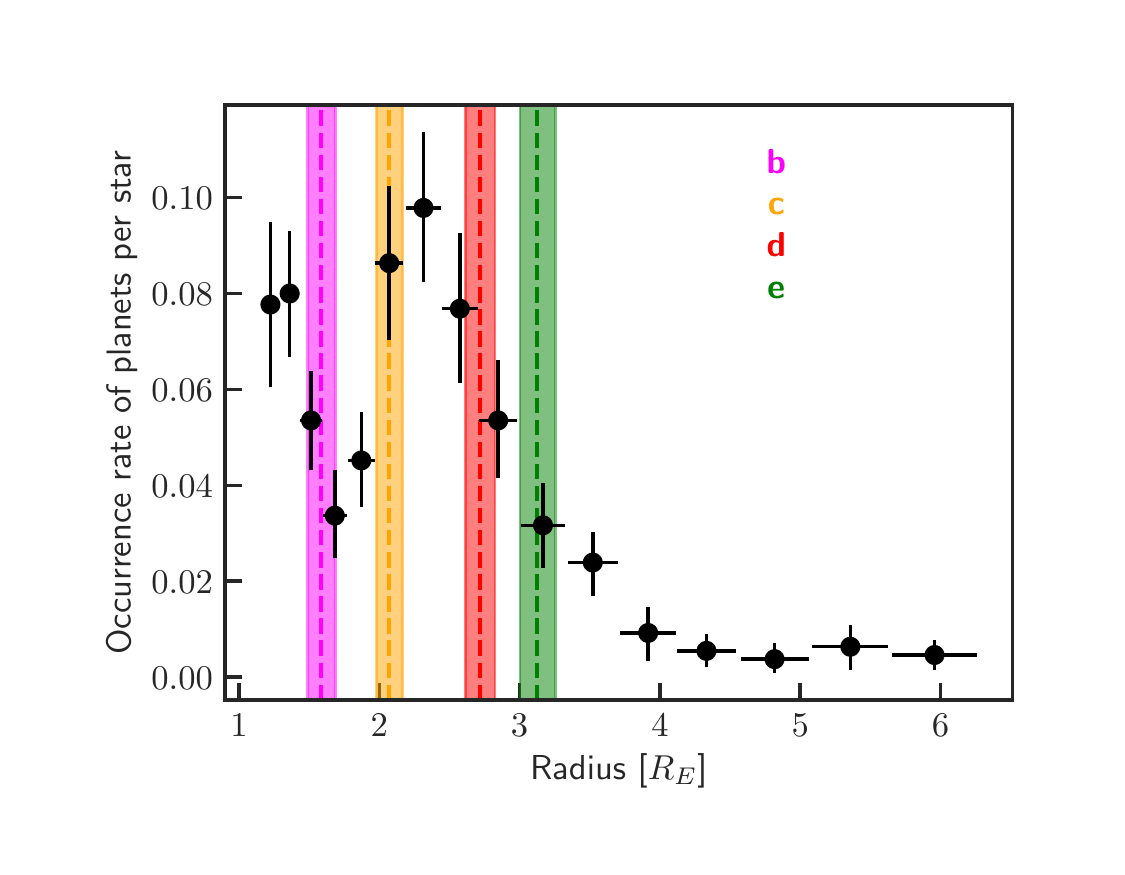}
    \caption{The radii of the planets transiting HD~108236 compared to the completeness-corrected occurrence rate of planets with orbital periods less than 100 days \citep{Fulton+2017}. The posterior median and 68\% credible interval of radii of the planets hosted by HD~108236 are highlighted with vertical lines and bands, respectively. Planet b falls within the radius valley \citep{Fulton+2017}.}
    \label{figr:histradi}
\end{figure}

Figure~\ref{figr:histradi} compares the inferred radii of the validated planets b, c, d, and e to the occurrence rate of planets as a function of planetary radius. Planet b is especially interesting for studies of photoevaporation, since its radius of \bRcompanionRearth{} $R_\oplus$ falls within a relatively uncommon radius range known as the radius valley \citep{Fulton+2017}. The radius valley is thought to be depleted due to photoevaporation caused by the stellar wind of the host star \citep{OwenWu2017}. However, the location of this radius valley has been shown to be a function of insolation flux \citep{Van-Eylen+2018}. Larger rocky planets can exist in more extremely irradiated environments. With an equilibrium temperature of \bTeq{}\,K, planet b is consistent with being part of the population of small, rocky planets just below the radius valley. In contrast, the planets c, d, and e are typical sub-Neptunes.

\subsection{Bright host}

\begin{figure*}
    \centering
    \includegraphics[trim=1cm 0.5cm 1cm 1cm, clip, width=0.49\textwidth]{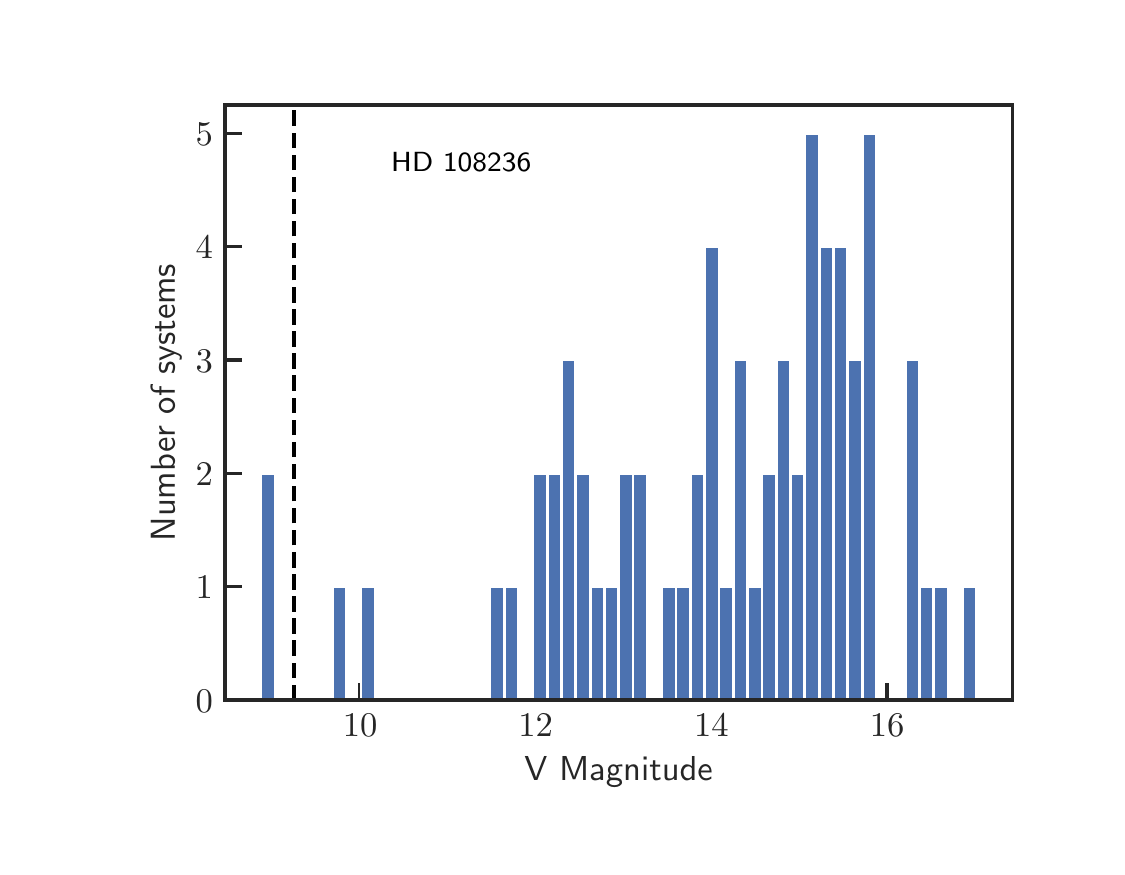}
    \includegraphics[trim=1cm 0.5cm 1cm 1cm, clip, width=0.49\textwidth]{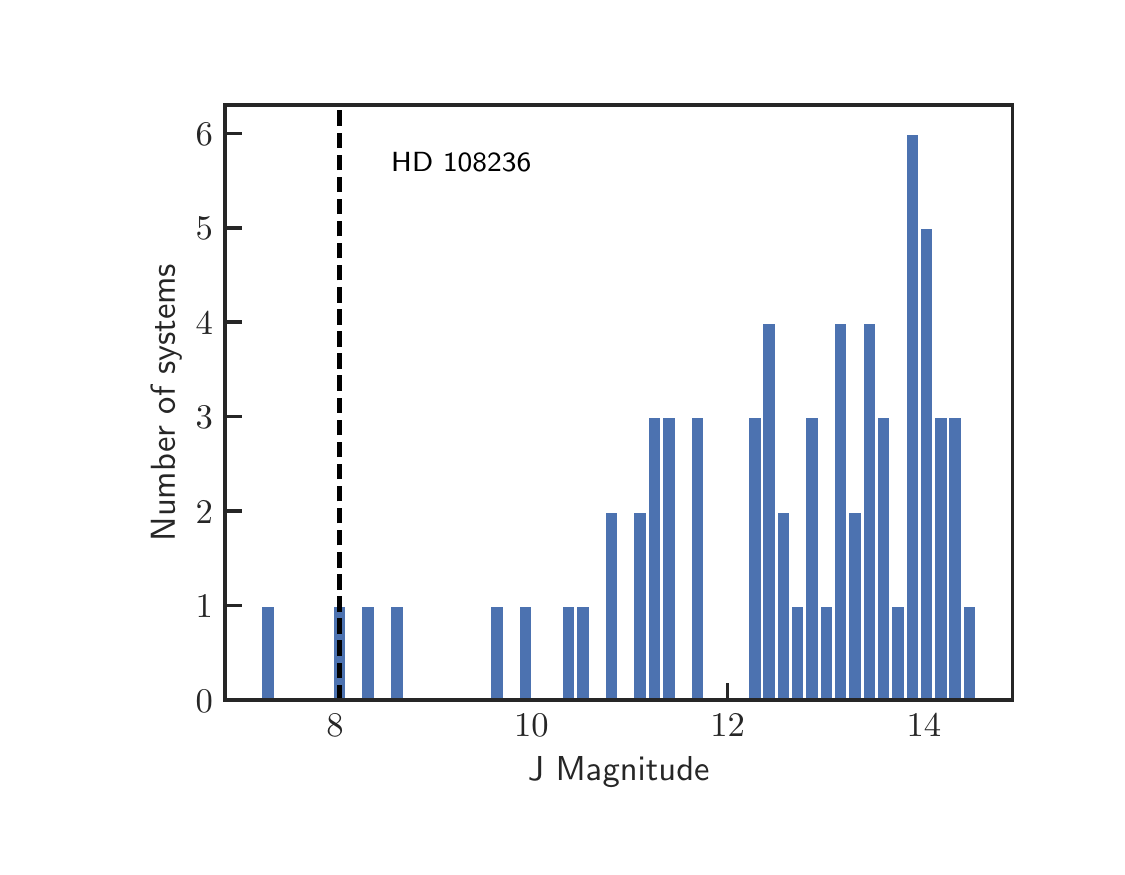} \\
    \includegraphics[trim=1cm 0.5cm 1cm 1cm, clip, width=0.49\textwidth]{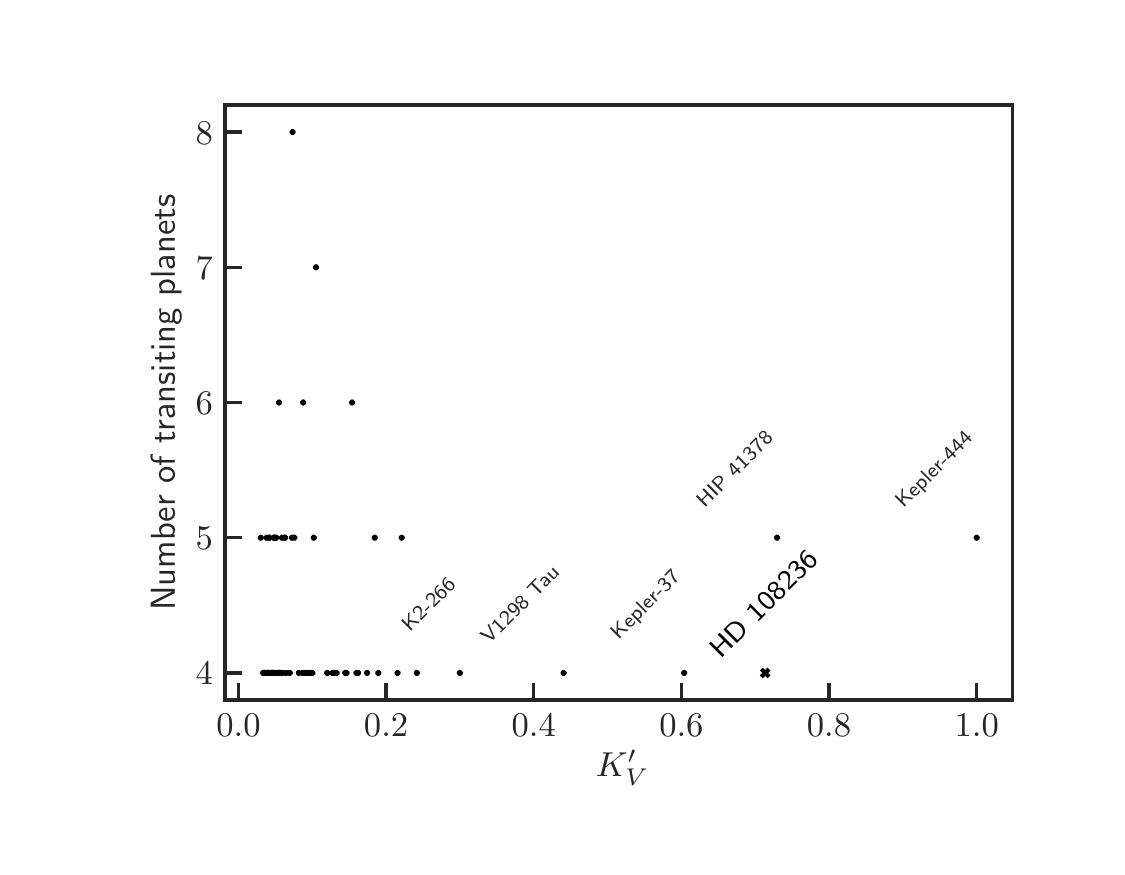}
    \includegraphics[trim=1cm 0.5cm 1cm 1cm, clip, width=0.49\textwidth]{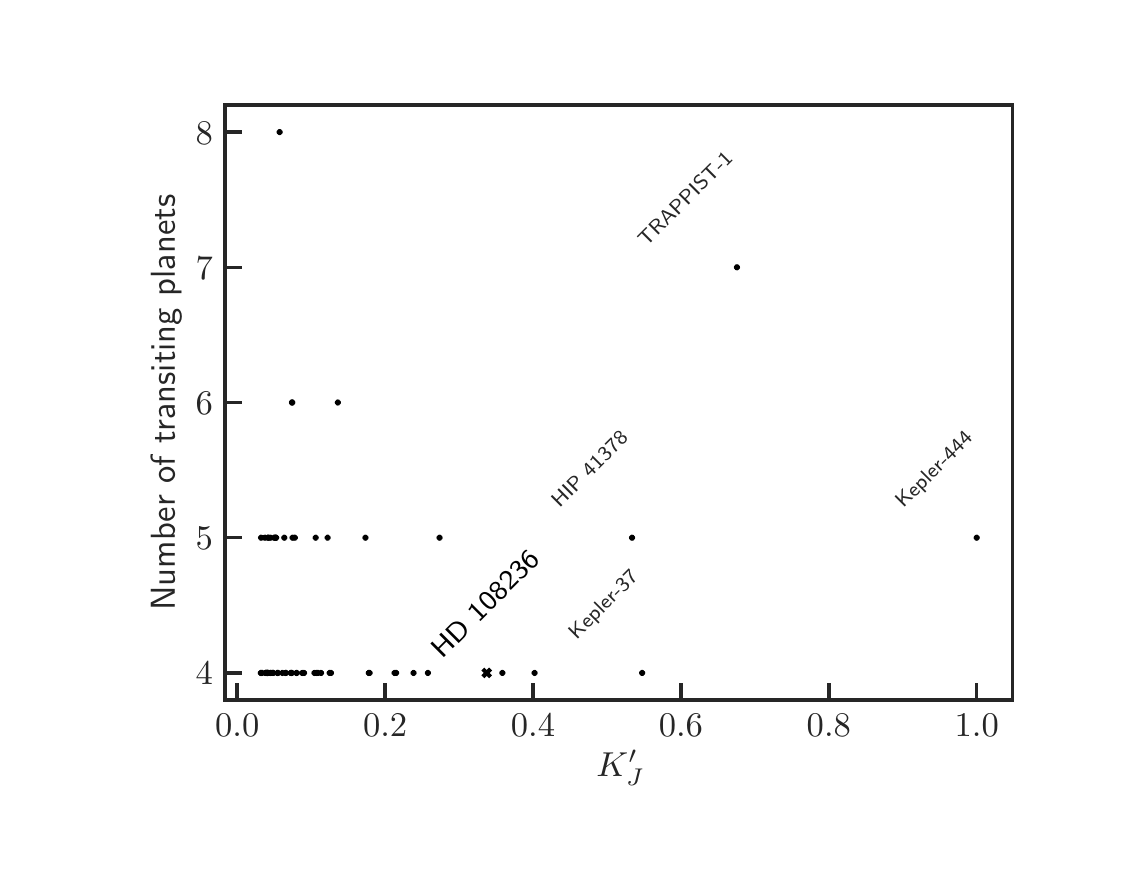}
    \caption{Comparison of HD~108236 to other systems with at least four transiting exoplanets. Top: histograms of the V (left) and J band (right) magnitudes  of systems that were previously known to host at least four transiting exoplanets. The magnitudes of HD~108236 are highlighted with dashed vertical black lines. Bottom: the radial velocity semi-amplitude (at fixed planet mass and orbital period) divided by the square root of the host star brightness in the V (left) and J (right) bands, denoted by $K^{\prime}_V$ and $K^{\prime}_J$. The x-axes are normalized such that the largest value is 1. We highlight the top 5 previously known systems retrieved from the NASA Exoplanet Archive. In the lower panel, the exoplanet labels are placed to the upper left of the corresponding points. HD~108236 is highlighted with crosses.}
    \label{figr:magt}
\end{figure*}

HD~108236 is one of the brightest stars that host four or more planets. As shown in the top row of Figure~\ref{figr:magt}, it is the third (behind Kepler 444 \citep{Campante+2015} and HIP 41378 \citep{Vanderburg+2016}) and the fourth brightest star (behind Kepler 444, HIP 41378, and Kepler 37 \citep{Barclay+2013}) in the V and J bands, respectively, that is known to host at least four planets. However, out of these, only Kepler 37 is a Sun-like star, making HD~108236 the brightest Sun-like star in the visual band to harbor at least four transiting planets. This property of HD~108236 makes it an interesting and accessible target from an observational point-of-view regarding future mass measurements, photometric follow-up and atmospheric characterization of its transiting planets.

The bottom row of Figure~\ref{figr:magt} also shows the radial velocity semi-amplitude (at fixed planet mass and orbital period) divided by the square root of the host star brightness in the V and J bands, respectively, which are denoted by $K^{\prime}_V$ and $K^{\prime}_J$. The x-axes are normalized so that the top target has the value of 1. Being a Sun-like star, HD~108236 falls to the 7$^{th}$ rank, when the comparison is made in the J band, since low-mass stars generate a larger radial velocity signal for a given companion.

\subsection{Mass measurement potential of the transiting planets}

The expected radial velocity semi-amplitudes of the four validated planets based on the predicted masses are in the range of 1.3--2.4 m s$^{-1}$. Given the brightness of the host star, this implies that the system has good potential for mass measurements in the near future. There are ongoing efforts to measure the masses of all validated transiting planets hosted by HD~108236.

Given the current absence of mass measurements of the planets, we use the probabilistic model of \citet{ChenKipping2017} in order to predict the masses of the validated planets. This model takes into account the measurement, sampling and intrinsic scatter of known planets in the mass-radius plane. As a result, the large uncertainties of the resulting mass predictions are dominated by this intrinsic system-to-system scatter and not by the posterior radius uncertainties of the planets validated in this work.

The masses of planets b, c, d, and e are predicted as $4\pm2$, $5\pm3$, $8\pm5$, and $10\pm6$ M$_\oplus$, respectively. Hence, planet b is likely a dense, rocky planet, whereas planets c, d, and e are sub-Neptunes with a hydrogen and helium envelope whose radius increases going from planet c to e. Atmospheric escape of volatiles is likely to be strongest for the innermost planet b, and should decrease towards the outermost planet e.

\subsection{Atmospheric characterization potential}
\label{sect:atmo}

Once the radius, mass and, hence, the bulk composition of a planet are determined, the next step in its characterization is the determination of its atmospheric properties. The available data on HD~108236 do not yet allow the atmospheric characterization of its planets. However, sub-Neptunes orbiting HD~108236 are favorable targets for near-future atmospheric characterization as we discuss below.

Given the expected launch of the James Webb Space Telescope (JWST), the Transmission Spectrum Metric (TSM) \citep{Kempton+2018},
\begin{equation}
    \text{TSM} \propto \dfrac{R^3_{\rm p} T_{\rm eq}}{M_{\rm p} R_\star^2},
    \label{equa:tsmm}
\end{equation}
ranks the relative SNR of different planets assuming observations made with the Near Infrared Imager and Slitless Spectrograph (NIRISS) \citep{Maszkiewicz2017} of JWST, assuming a cloud-free, hydrogen-dominated atmosphere. 

The largest uncertainty in predicting the TSMs of the planets orbiting HD~108236 arises from the current unavailability of their mass measurements. We use the predicted masses of planet b, c, d, and e in Equation~\ref{equa:tsmm} to obtain \emph{preliminary} estimates of their TSMs. Based on the brightness of the host star, it is expected that the masses of all validated planets will be measured to better than 40\%. Therefore, comparing the TSMs of the validated planets to those of all known sub-Neptunes retrieved from the NASA Exoplanet Archive\footnote{\url{https://exoplanetarchive.ipac.caltech.edu/}} Planetary Systems Composite Data with mass measurement uncertainties better than 5$\sigma$, we find that the sub-Neptunes HD~108236\,c, HD~108236\,d, and HD~108236\,e fall among the top 20. The super-Earth (planet b) is not included in this TSM ranking, because it is not expected to have a hydrogen-dominated atmosphere. We once again emphasize that these rankings are based on the \emph{predicted} masses and the actual rankings will depend on the mass measurements of the planets.

\begin{figure}[!htbp]
    \centering
    \includegraphics[trim=0.5cm 0cm 0.5cm 0.5cm, clip, width=0.49\textwidth]{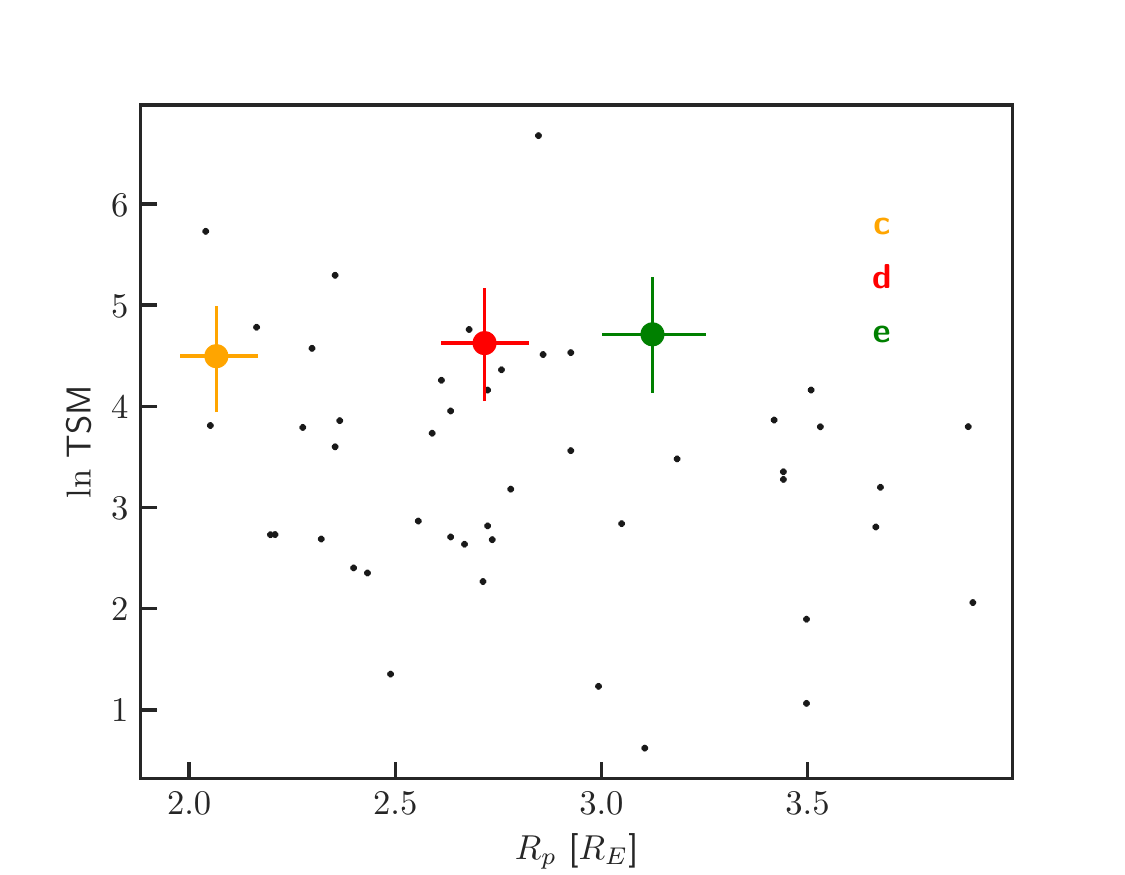}
    \caption{The logarithm of TSM vs. radius distribution of the sub-Neptunes orbiting HD~108236 and known transiting planets with mass measurements better than 5$\sigma$ as retrieved from the NASA Exoplanet Archive. Planets c, d, and e of HD~108236 are among the top 20 known sub-Neptunes when ranked with respect to their TSMs.}
    \label{figr:tsmm}
\end{figure}

The logarithms of the relative TSMs of the planets are plotted against their radii in Figure~\ref{figr:tsmm}, along with those of the known exoplanets (black points) retrieved from the NASA Exoplanet Archive, where the overall normalizations of the TSMs is arbitrary. We only show those known planets that have a measured mass with an uncertainty better than 40\%. The three sub-Neptunes of the HD~108236 system are found to be favorable targets for comparative characterization of sub-Neptune atmospheres. 

It is worth noting that the TSM ranking of the HD~108236 sub-Neptunes improves with decreasing equilibrium temperature, despite the fact that lowering the temperature acts to reduce the pressure scale height. 

As can be seen in Equation~\ref{equa:tsmm}, the TSM is proportional to the third power of $R_p$, while inversely proportional to $M_p$. Although it also scales with $M_p$, the $R_p$ dependence of $M_p$ is weaker than $R_p^3$. Therefore, the TSM is more sensitive to an increase in planetary radius than a drop in equilibrium temperature. In the HD~108236 system, the radii of the planets c, d, and e increase with decreasing equilibrium temperature. As a result, the predicted TSM increases from planet c to e.

Furthermore, although HD~108236 is a relatively bright target, its brightness is below the limiting magnitude of NIRISS/JWST (J magnitude of $\sim$ 7) \citep{Beichman+2014}, making it an appealing transmission spectroscopy target for the instrument.

\begin{figure}[!htbp]
    \centering
    \includegraphics[trim=0cm 0cm 0.5cm 0.5cm, clip, width=0.49\textwidth]{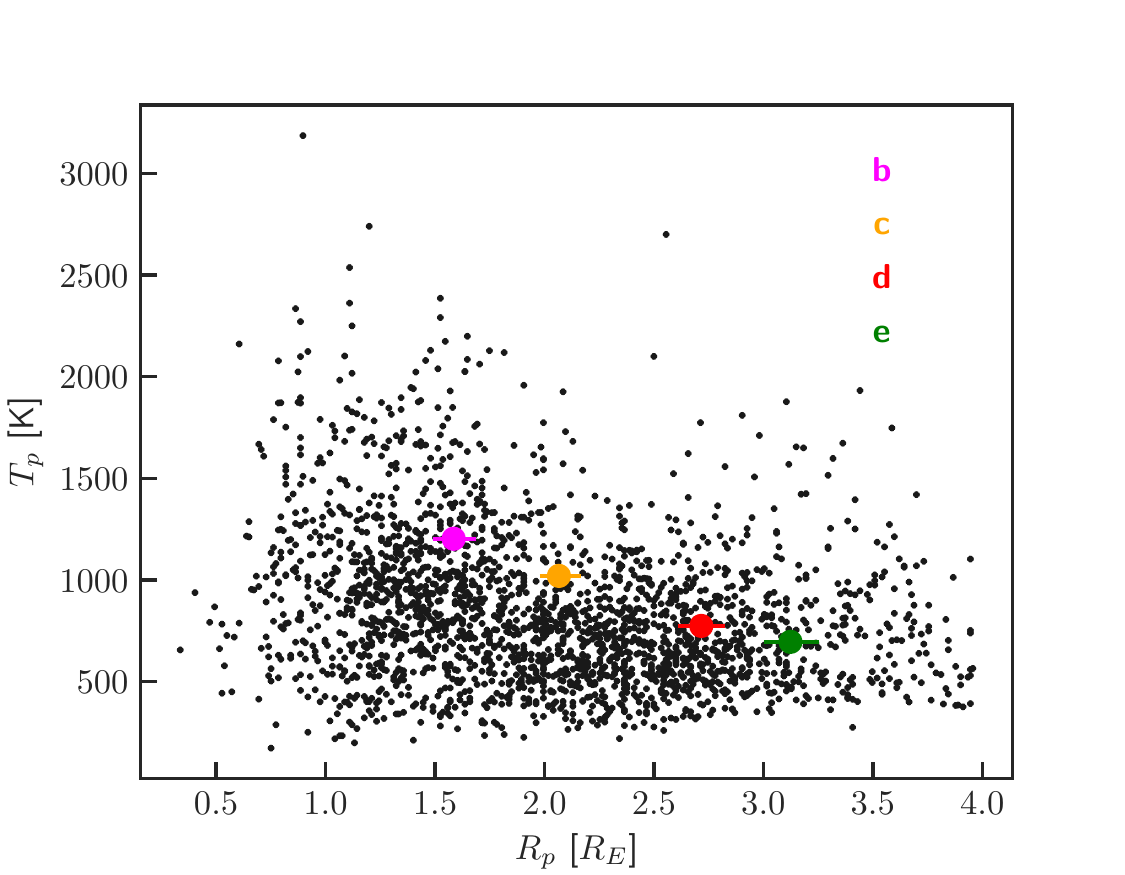}
    \caption{The equilibrium temperatures and radii of known planets retrieved from the NASA Exoplanet Archive, shown with black points. Planets orbiting HD~108236 are highlighted, which span a broad and representative range of radius and equilibrium temperature.}
    \label{figr:raditmpt}
\end{figure}

We also note that planets orbiting HD~108236 span a broad range of radius and equilibrium temperature. Figure~\ref{figr:raditmpt} shows the distribution of radii and equilibrium temperatures of known planets retrieved from the NASA Exoplanet Archive and those of the planets orbiting HD~108236. The wide range of radii and equilibrium temperatures manifested by the planets allows controlled experiments of how stellar insolation affects the photoevaporation of the volatile envelopes of the orbiting planets by controlling for the stellar type and evolution history\citep{OwenCampos-Estrada2020}.

\subsection{Dynamics}

In a multiplanetary system, the displacement from a mean motion resonance (MMR)
\begin{equation}
\Delta = \frac{P'}{P}\frac{j-k}{j}-1, 
\end{equation}
of adjacent planet pairs characterizes the proximity of the pair to a MMR, where $P'$ and $P$ are the orbital periods of the outer and inner planets, $j$ is the nearest integer to the orbital period ratio, and $k$ is the order of the nearest MMR. Proximity to an MMR results in TTVs with a coherence time scale (i.e., super-period) of $P_{\rm ttv}$ such that
\begin{equation}
\frac{1}{P_{\rm ttv}} = \left|\frac{j-k}{P}-\frac{j}{P'}\right|.
\end{equation}
The HD~108236 system consists of closely packed planets. However, no pair of the validated planets is on an MMR. The proximities and super-periods of the known adjacent pairs in the HD~108236 system are shown in Table~\ref{tabl:dyna}.

\begin{table}[h!]
    \begin{center}
        \begin{tabular}{|c|c|c|c|c|c|c|c|}
            \hline
            Pair & $P'/P$ & j:j-k & $\Delta$ & P$_{\rm ttv}$ [day] \\
            \hline
             b,c & 1.63473 & 5:3 & -0.01916 &  64.75626 \\ 
             c,d & 2.28506 & 9:4 & 0.01558 & 101.08835 \\ 
             d,e & 1.37870 & 4:3 & 0.03403 & 143.61021 \\ 
            \hline
        \end{tabular}
        \caption{Proximities to MMRs of adjacent planet pairs in the HD~108236 system. The second and third columns list the orbit period ratios and nearest MMR, while the fourth and fifth columns estimate the proximity to resonance and the coherence period respectively. The outer pair is near a first order resonance where we noted the estimated TTV amplitude in the last two columns, as described in the text.}
    \end{center}
    \label{tabl:dyna}
\end{table}

For the first order interaction between a pair, where $k=1$, the amplitude of the TTVs, $V$ and $V'$, can be estimated using the analytical solution \citep{Lithwick+2012}
\begin{align}
    V &= P \frac{\mu'}{\pi j^{2/3} (j-1)^{1/3}\Delta}\left(-f - \frac{3}{2} \frac{Z_{\rm free}^{*}}{\Delta} \right), \\
    V' &= P' \frac{\mu}{\pi j \Delta}\left(-g + \frac{3}{2} \frac{Z_{\rm free}^{*}}{\Delta} \right),
\end{align}
where $f$ and $g$ are coefficients, $\mu$ and $\mu'$ are the masses of the planets normalized by that of the host star, and $Z_{\rm free}^{*}$ is the conjugate of the complex sum of eccentricity vectors.

No planet pairs in the HD~108236 system are in or near a strong MMR, precluding the generation of large resonant TTVs. However, non-resonant (chopping) TTVs with small amplitudes induced by synodic interactions, are possible. Assuming circular orbits and using the predicted masses yield a TTV of $\sim$ 5 minutes for both planet d and e. We also confirmed this analytical prediction using an N-body dynamical simulation \citep{Lissauer+2011} of HD~108236 with a length of 5000 days. We note that the planets could have higher TTVs when the circular orbit assumption is relaxed. Hence, with sufficient transit timing precision, planets d and e are likely to be amenable to mass measurements via TTV observations enabled by long-term transit photometry follow-up \citep{DeckAgol2015}.

\paragraph{Potential 3-body resonances due to a hypothetical planet x}

The orbital gaps between planet b and c and between planet c and d are large enough for low mass planets to exist on stable orbits, as is common among multiplanetary systems discovered by the Kepler telescope. There are many adjacent pairs in the Kepler data set close to the 3:2 MMR, which invokes the possibility of a missing planet in the apparent 9:4 near resonant gap between the middle pair of HD~108236. While the parameter space for such missing planets is fairly large, we note that resonant chains of 3 bodies, as is present in systems like TRAPPIST-1 \citep{Gillon+2017} and Kepler-80 \citep{Xie2013}, could be present in HD~108236 due to yet-undetected planets. This undetected planet x could either have a period of $P_{\rm x}$  = 9.364 days, which would satisfy
\begin{equation}
0 \approx 2n_{\rm c}-5n_{\rm x} +3n_{\rm d},
\end{equation}
where $n_{\rm x}$ is the orbital frequency of the hypothetical planet, or a period of P$_{\rm x} $ = 9.150 days, which would satisfy
\begin{equation}
0 \approx n_{\rm x}-3n_{\rm d} +2n_{\rm e}.
\end{equation}
The resulting 3:2 resonance between this hypothetical planet x and planet d would result in additional TTVs.

To search for evidence of such an additional planet in the TESS data, we used \texttt{allesfitter}'s interface to remove the remnant stellar variability from the PDC light curve using a cubic spline and recursive sigma clipping via \texttt{wotan} \citep{Hippke+2019}. Then, we ran a \texttt{TLS} search \citep{HippkeHeller2019} on this flattened light curve. We recovered all four transiting planets b,c, d, and e. We also detected several additional periodic transit-like signals above an SNR threshold of 5. The most statistically significant of these detections has an epoch of 2458570.6781 BJD, period of $10.9113$\,days, transit depth of $0.23$\,ppt, SNR of 8.0, signal detection efficiency (SDE) of 6.9, and false alarm probability of 0.01. We therefore present this as a tentative fifth planet candidate in the HD~108236 system. However, given the large false positive probability and its dependence on the detrending method, we concluded that instrumental origin cannot be ruled out for this planet candidate. In particular, the stellar density consistent with the transits of this candidate is $0.4\pm0.3$ g cm$^{-3}$, which is inconsistent with the stellar density ($1.9$ g cm$^{-3}$) inferred in Section~\ref{sect:star}. This implies that the candidate is likely due to systematics. Given the larger false positive probabilities of the other \texttt{TLS} detections (i.e., larger than 0.01), we discarded them as likely due to systematics in the TESS data.

\paragraph{TTV analysis of TESS transits}

\begin{figure}[!htbp]
    \centering
    \includegraphics[trim=0.5cm 1cm 0.5cm 1cm, clip, width=0.49\textwidth]{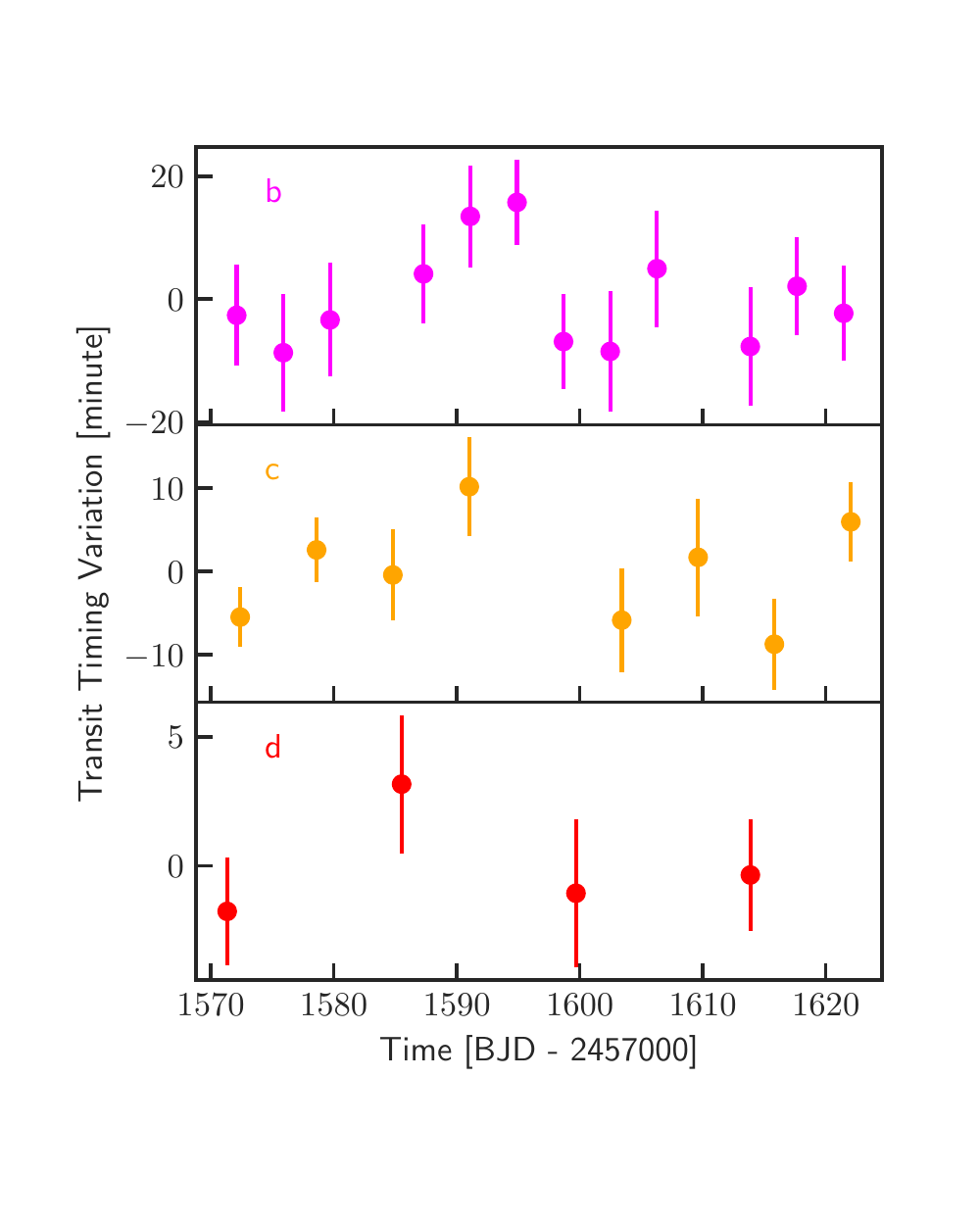}
    \caption{The measured TTVs of the discovered planets in the HD~108236 system. The measured mid-transit times are consistent with a linear ephemeris model. No TTV for planet e was measured, since only two transits were observed.}
    \label{figr:ttvr}
\end{figure}

In order to infer the TTVs consistent with the TESS data, we performed a light curve analysis independent of that discussed in Section~\ref{sect:modl} using \texttt{exoplanet} \citep{Foreman-Mackey+2020} by relaxing the assumption of a linear ephemeris. The resulting TTVs are shown in Figure~\ref{figr:ttvr}. Table~\ref{tabl:ttvrmeas} also tabulates the mid-transit times of the transits detected in the TESS data. We did not detect any significant TTVs given the temporal baseline and timing precision of the transits observed by TESS. Nevertheless, using these TTVs, we were able to constrain the mass of planet e to be lower than 31 M$_\oplus$ at 2$\sigma$ via the dynamical simulation, which is consistent with the mass predicted via \citet{ChenKipping2017}.

\begin{table}[h!]
    \caption{Measured mid-transit times of planets b, c, and d in the TESS data. All times are provided in BJD after subtracting 2,457,000.}
    \begin{center}
        \begin{tabular}{cc}
            Mid-transit time [BJD - 2,457,000] & 1$\sigma$ uncertainty [days] \\
            \hline
            \hline
            Planet b \\
            \hline
            1572.107037	& 0.006751046 \\
            1575.898507	& 0.007962894 \\
            1579.697924	& 0.007157883 \\
            1587.294548	& 0.00576889 \\
            1591.096759	& 0.005991691 \\
            1594.894048	& 0.00481626 \\
            1598.673998	& 0.005489018 \\
            1602.468591	& 0.007256515 \\
            1606.273666	& 0.007104524 \\
            1613.856271	& 0.007697341 \\
            1617.658793	& 0.006202734 \\
            1621.451437	& 0.00614042 \\
            \hline
            Planet c \\
            \hline
            1572.391729	& 0.002815299 \\
            1578.601024	& 0.002967442 \\
            1584.802628	& 0.004321249 \\
            1591.013683	& 0.004541912 \\
            1603.409944	& 0.004748817 \\
            1609.618876	& 0.005754455 \\
            1615.815326	& 0.004564704 \\
            1622.029226	& 0.003369172 \\
            \hline
            Planet d \\
            \hline
            1571.335310	& 0.00213619 \\
            1585.514907	& 0.002414469 \\
            1599.688154	& 0.002331228 \\
            1613.864821	& 0.002721803 \\
            \hline
        \end{tabular}
    \end{center}
    \label{tabl:ttvrmeas}
\end{table}

\paragraph{Stability}

\begin{figure*}[!htbp]
  \begin{center}
    \includegraphics[trim=1cm 0.5cm 2cm 0cm, clip, width=\textwidth]{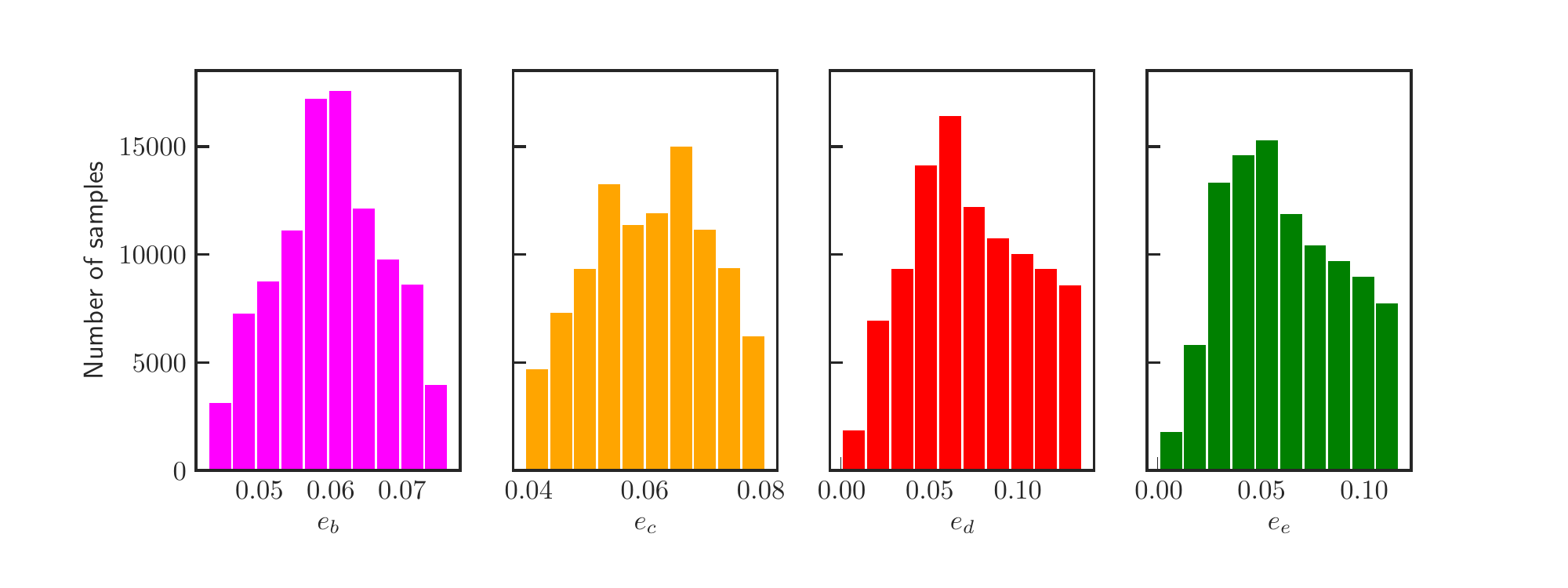}
  \end{center}
  \caption{Histograms of the eccentricities inferred from the dynamical stability simulation. The system retains orbital integrity throughout the $10^7$ year simulation time.}
  \label{figr:histecce}
\end{figure*}

To further test the dynamical integrity of the system, we conducted N-body integrations using the Mercury Integrator Package \citep{Chambers1999}. Our method is similar to that adopted by \citep{Kane2015, Kane2019} in the study of compact planetary systems discovered by Kepler. The innermost planet of our system has an orbital period of $\sim$3.8~days. To ensure perturbative accuracy, we therefore used a conservative time step for the simulations of 0.1 days, which is $\sim1/40$ of the period of the innermost planet. We ran the simulation for $10^7$ years, equivalent to $\sim 10^9$ orbits of the innermost planet. For the masses of the planets b, c, d, and e, we assumed fiducial values of 3.5, 4.7, 7.2, and 11.1 M$_\oplus$. The results of the simulation are represented in Figure~\ref{figr:histecce} by showing the histogram of the eccentricities of the four planets for the entire simulation. The results show that the system is dynamically stable, even considering the non-zero eccentricities for such a compact system. However, there is significant transfer of angular momentum that occurs between the planets with time. The two innermost planets have eccentricities that oscillate between 0 and $\sim 0.13$, which can result in substantial changes in the climate of the atmospheres \citep{KaneTorres2017, WayGeograkarakos2017}, known as Milankovitch cycles \citep{Spiegel+2010}. The two outermost planets, d and e, remain near their starting eccentricities and so are largely unperturbed through the orbital evolution.

\section{Discussion and Conclusion}
\label{sect:disc}

Systems with multiple planets provide a test bed for models of planet formation, evolution and orbital migration. Roughly one-third of the planetary systems discovered by the Kepler telescope are multiplanetary \citep{Borucki+2011}. The inferred valley in the radius distribution of known, small planets \citep{Fulton+2017} is possibly due to the photoevaporation of volatile gases on close-in planets or core-powered mass loss \citep{Ginzburg2018}. These processes can leave behind a rocky core and a small (less than 2 $R_\oplus$) radius, while the unaffected population constitute gas giants with radii larger than 2 $R_\oplus$. Furthermore, if photoevaporation is indeed the mechanism that causes the radius valley, then adjacent planets in multiplanetary systems should have similar radii, since they have had similar irradiation histories. The planets of HD~108236 are consistent with this model, since the radius ratios of adjacent planets are 1.3, 1.3, and 1.1, respectively.

Regarding its coplanar and compact nature, the orbital architecture of the HD~108236 multiplanetary system is also consistent with those of the multiplanetary systems discovered by the Kepler telescope. The CKS sample of exoplanets exhibited a correlation between the size and spacing of the planets \citep{Weiss+2018a, FangMargot2013}, which is also demonstrated in the HD~108236 system. That is, adjacent planets are found to have similar sizes and their period ratios are correlated. Furthermore, in the CKS sample, the period ratio of adjacent planets were observed to cluster just above 1.2, with very few period ratios of adjacent planets below 1.2. This can either be due to in-situ formation at these period ratios or due to subsequent orbital migration. In either case, it was determined that this period ratio defines a stability region \citep{Weiss+2018a}, as pairs with a period ratio smaller than 1.2 become dynamically unstable due to Hill or Lagrange instability. With period ratios of \cperiodoverbperiod{}, \dperiodovercperiod{}, and \eperiodoverdperiod{}, planets discovered in this work also respect this dynamical constraint.

In short, HD~108236 offers an excellent laboratory for studying planet formation and evolution as well as atmospheric characterization while controlling for the stellar type and age. The sub-Neptunes HD~108236\,c, HD~108236\,d, and HD~108236\,e will be favorable targets for atmospheric characterization via transmission spectroscopy with the JWST and HST. The brightness of the host, its similarity to the Sun and the potentially yet-unknown outer companions makes the system a high-priority target for characterization. The target will be reobserved in the extended mission of TESS during Cycle 3, Sector 37 (UT 2 April 2021 to UT 28 April 2021, which will enable improved TTV measurements and searches for new transiting planets in the system. HD~108236 will also be among the targets observed by CHEOPS for improved radius characterization.

\section*{Acknowledgments}

This paper includes data collected by the TESS mission, which are publicly available from the Mikulski Archive for Space Telescopes (MAST). Funding for the TESS mission is provided by NASA's Science Mission directorate. We acknowledge the use of public TESS Alert data from pipelines at the TESS Science Office and at the TESS Science Processing Operations Center. This research has also made use of the Exoplanet Follow-up Observation Program website, which is operated by the California Institute of Technology, under contract with the National Aeronautics and Space Administration under the Exoplanet Exploration Program. Resources supporting this work were provided by the NASA High-End Computing (HEC) Program through the NASA Advanced Supercomputing (NAS) Division at Ames Research Center for the production of the SPOC data products.

The MEarth Team gratefully acknowledges funding from the David and Lucile Packard Fellowship for Science and Engineering (awarded to D.C.). This material is based upon work supported by the National Science Foundation under grants AST-0807690, AST-1109468, AST-1004488 (Alan T. Waterman Award), and AST-1616624. This work is made possible by a grant from the John Templeton Foundation. The opinions expressed in this publication are those of the authors and do not necessarily reflect the views of the John Templeton Foundation. This material is based upon work supported by the National Aeronautics and Space Administration under Grant No. 80NSSC18K0476 issued through the XRP Program.

Some of the Observations in the paper made use of the High-Resolution Imaging instrument Zorro. Zorro was funded by the NASA Exoplanet Exploration Program and built at the NASA Ames Research Center by Steve B. Howell, Nic Scott, Elliott P. Horch, and Emmett Quigley. Zorro was mounted on the Gemini South telescope of the international Gemini Observatory, a program of NSF’s OIR Lab, which is managed by the Association of Universities for Research in Astronomy (AURA) under a cooperative agreement with the National Science Foundation. on behalf of the Gemini partnership: the National Science Foundation (United States), National Research Council (Canada), Agencia Nacional de Investigación y Desarrollo (Chile), Ministerio de Ciencia, Tecnología e Innovaci\'{o}n (Argentina), Ministério da Ciência, Tecnologia, Inova\c{c}\~{o}es e Comunica\c{c}\~{o}es (Brazil), and Korea Astronomy and Space Science Institute (Republic of Korea).

This work makes use of observations from the LCOGT network.

Based in part on observations obtained at the Southern Astrophysical Research (SOAR) telescope, which is a joint project of the Ministério da Ciência, Tecnologia e Inovações do Brasil (MCTI/LNA), the US National Science Foundation’s NOIRLab, the University of North Carolina at Chapel Hill (UNC), and Michigan State University (MSU).

Support for this work was provided by NASA through grant 18-XRP18\_2-0048.

TD acknowledges support from MIT's Kavli Institute as a Kavli postdoctoral fellow.
MNG acknowledges support from MIT's Kavli Institute as a Torres postdoctoral fellow.

Contributions from KP and JW were made through the Harvard-MIT Science Research Mentoring Program \citep{Graur2018}, led by Clara Sousa-Silva through the 51 Pegasi b Fellowship, and Or Graur. Support for this program is provided by the National Science Foundation under award AST-1602595, City of Cambridge, the John G. Wolbach Library, Cambridge Rotary, Heising-Simons Foundation, and generous individuals.

We thank Edward Bryant and the NGTS \citep{Wheatley+2018} team for their HD~108236 observation attempts.

\textit{Facilities}:
{TESS, LCOGT, Magellan II, SMARTS, Gemini, SOAR}

\textit{Software}:
\texttt{python} \citep{Rossum1995},
\texttt{matplotlib} \citep{Hunter2007},
\texttt{seaborn}
(\url{https://seaborn.pydata.org/index.html}),
\texttt{numpy} \citep{vanderWalt2011},
\texttt{scipy} \citep{Jones2001}, 
\texttt{allesfitter} \citep[][and in prep.]{GuentherDaylan2019a, GuentherDaylan2019b},
\texttt{ellc} \citep{Maxted2016},
\texttt{EXOFASTv2} \citep{Eastman+2019},
\texttt{emcee} \citep{Foreman-Mackey2013},
\texttt{celerite} \citep{Foreman-Mackey2017},
\texttt{corner} \citep{Foreman-Mackey2016}.
\texttt{dynesty} \citep{Speagle2019},
\texttt{AstroImageJ} \citep{Collins+2017},
\texttt{Tapir} \citep{Jensen2013},
\textsf{exoplanet} \citep{Foreman-Mackey+2020},
\textsf{Transit Least Squares} \citep{HippkeHeller2019}, 
\textsf{astroquery} \citep{Ginsburg+2019}, 
\textsf{Lightkurve} \citep{Cardoso+2018}, 
\textsf{pymc3} \citep{Salvatier+2016},

\begin{table}[!htbp]
	\centering
	\caption{Posterior of the fitting nuisance parameters.}
	\begin{tabular}{cccc}
parameter & value & unit & fit/fixed \\ 
\hline 
\hline 
$D_\mathrm{0; TESS}$ & $0.0$ & & fixed \\ 
$q_{1; \mathrm{TESS}}$ & $0.23_{-0.11}^{+0.19}$ & & fit \\ 
$q_{2; \mathrm{TESS}}$ & $0.43_{-0.29}^{+0.36}$ & & fit \\ 
$\log{\sigma_\mathrm{TESS}}$ & $-7.4845\pm0.0090$ & $\log{ \mathrm{rel. flux.} }$& fit \\ 
$\log{\sigma_{\rm GP;\mathrm{TESS}}}$ & $-8.56\pm0.13$ & & fit \\ 
$\log{\rho_{\rm GP;\mathrm{TESS}}}$ & $-1.27\pm0.28$ & & fit \\ 
        \end{tabular}
   \label{tabl:parafittnuis}
\end{table}

\begin{table}[!htbp] 
	\centering
	\caption{Posterior of the fitting parameters for planets b and c.}
	\begin{tabular}{cccc}
parameter & value & unit & fit/fixed \\ 
\hline 
\hline 
$R_{\rm b} / R_\star$ & $0.01638\pm0.00095$ & & fit \\ 
$(R_\star + R_{\rm b}) / a_b$ & $0.0895_{-0.0025}^{+0.0028}$ & & fit \\ 
$\cos{i_{\rm b}}$ & $0.037_{-0.022}^{+0.015}$ & & fit \\ 
$T_{0;\rm b}$ & $2458572.1128_{-0.0036}^{+0.0031}$ & $\mathrm{BJD}$& fit \\ 
$P_{\rm b}$ & $3.79523_{-0.00044}^{+0.00047}$ & $\mathrm{d}$& fit \\ 
$\sqrt{e_{\rm b}} \cos{\omega_{\rm b}}$ & $-0.00\pm0.50$ & & fit \\ 
$\sqrt{e_{\rm b}} \sin{\omega_{\rm b}}$ & $-0.03_{-0.31}^{+0.27}$ & & fit \\ 
$R_c / R_\star$ & $0.02134_{-0.00083}^{+0.00094}$ & & fit \\ 
$(R_\star + R_{\rm c}) / a_{\rm c}$ & $0.0647_{-0.0019}^{+0.0021}$ & & fit \\ 
$\cos{i_{\rm c}}$ & $0.022_{-0.014}^{+0.013}$ & & fit \\ 
$T_{0;\rm c}$ & $2458572.3949_{-0.0020}^{+0.0025}$ & $\mathrm{BJD}$& fit \\ 
$P_c$ & $6.20370_{-0.00052}^{+0.00064}$ & $\mathrm{\rm d}$& fit \\ 
$\sqrt{e_{\rm c}} \cos{\omega_{\rm c}}$ & $-0.01\pm0.49$ & & fit \\ 
$\sqrt{e_{\rm c}} \sin{\omega_{\rm c}}$ & $-0.11_{-0.29}^{+0.23}$ & & fit \\ 
        \end{tabular}
   \label{tabl:parafittseco}
\end{table}

\begin{table}[!htbp]
	\centering
	\caption{Posterior of the fitting parameters for planets d and e.}
	\begin{tabular}{cccc}
parameter & value & unit & fit/fixed \\ 
\hline 
\hline 
$R_{\rm d} / R_\star$ & $0.02805\pm0.00095$ & & fit \\ 
$(R_\star + R_{\rm d}) / a_{\rm d}$ & $0.0375_{-0.0010}^{+0.0012}$ & & fit \\ 
$\cos{i_{\rm d}}$ & $0.0136_{-0.0078}^{+0.0065}$ & & fit \\ 
$T_{0;\rm d}$ & $2458571.3368_{-0.0013}^{+0.0015}$ & $\mathrm{BJD}$& fit \\ 
$P_{\rm d}$ & $14.17555_{-0.0011}^{+0.00099}$ & $\mathrm{d}$& fit \\ 
$\sqrt{e_{\rm d}} \cos{\omega_{\rm d}}$ & $-0.03_{-0.48}^{+0.51}$ & & fit \\ 
$\sqrt{e_{\rm d}} \sin{\omega_{\rm d}}$ & $-0.04_{-0.27}^{+0.21}$ & & fit \\ 
$R_{\rm e} / R_\star$ & $0.0323_{-0.0011}^{+0.0012}$ & & fit \\ 
$(R_\star + R_{\rm e} / a_{\rm e}$ & $0.03043_{-0.00089}^{+0.00100}$ & & fit \\ 
$\cos{i_{\rm e}}$ & $0.0118_{-0.0073}^{+0.0052}$ & & fit \\ 
$T_{0;\rm e}$ & $2458586.5677\pm0.0014$ & $\mathrm{BJD}$& fit \\ 
$P_e$ & $19.5917_{-0.0020}^{+0.0022}$ & $\mathrm{d}$& fit \\ 
$\sqrt{e_{\rm e}} \cos{\omega_{\rm e}}$ & $0.01_{-0.54}^{+0.50}$ & & fit \\ 
$\sqrt{e_{\rm e}} \sin{\omega_{\rm e}}$ & $0.02_{-0.29}^{+0.23}$ & & fit \\ 
        \end{tabular}
   \label{tabl:parafittfrst}
\end{table}

\begin{table}[!htbp] 
	\centering
	\caption{Posterior of the inferred parameters for planets b and c.}
	\begin{tabular}{cc}
Property & Value \\ 
\hline 
\hline 
$R_\star/a_\mathrm{b}$ & $0.0881_{-0.0025}^{+0.0027}$  \\ 
$a_\mathrm{b}/R_\star$ & $11.35\pm0.34$  \\ 
$R_\mathrm{b}/a_\mathrm{b}$ & $0.001443_{-0.000092}^{+0.000100}$  \\ 
$R_\mathrm{b}$ ($\mathrm{R_{\oplus}}$) & $1.586\pm0.098$  \\ 
$R_\mathrm{b}$ ($\mathrm{R_{jup}}$) & $0.1415\pm0.0087$  \\ 
$a_\mathrm{b}$ ($\mathrm{R_{\odot}}$) & $10.08\pm0.36$  \\ 
$a_\mathrm{b}$ (AU) & $0.0469\pm0.0017$  \\ 
$i_\mathrm{b}$ (deg) & $87.88_{-0.87}^{+1.3}$  \\ 
$e_\mathrm{b}$ & $0.20_{-0.14}^{+0.30}$  \\ 
$w_\mathrm{b}$ (deg) & $190\pm140$  \\ 
$b_\mathrm{tra;b}$ & $0.38\pm0.24$  \\ 
$T_\mathrm{tot;b}$ (h) & $2.30_{-0.11}^{+0.16}$  \\ 
$T_\mathrm{full;b}$ (h) & $2.20_{-0.12}^{+0.16}$  \\ 
$\rho_\mathrm{\star;b}$ (cgs) & $1.92\pm0.17$  \\ 
$T_\mathrm{eq;b}$ (K) & $1099_{-18}^{+19}$  \\
$\delta_\mathrm{tr; b; TESS}$ (ppt) & $0.302\pm0.031$  \\ 
$P_\mathrm{b} / P_\mathrm{c}$ & $0.611768_{-0.000098}^{+0.000092}$  \\ 
$P_\mathrm{b} / P_\mathrm{d}$ & $0.267731\pm0.000038$  \\ 
$P_\mathrm{b} / P_\mathrm{e}$ & $0.193716\pm0.000031$  \\ 
$R_\star/a_\mathrm{c}$ & $0.0634_{-0.0018}^{+0.0020}$  \\ 
$a_\mathrm{c}/R_\star$ & $15.78\pm0.49$  \\ 
$R_\mathrm{c}/a_\mathrm{c}$ & $0.001354_{-0.000067}^{+0.000076}$  \\ 
$R_\mathrm{c}$ ($\mathrm{R_{\oplus}}$) & $2.068_{-0.091}^{+0.10}$  \\ 
$R_\mathrm{c}$ ($\mathrm{R_{jup}}$) & $0.1845_{-0.0081}^{+0.0089}$  \\ 
$a_\mathrm{c}$ ($\mathrm{R_{\odot}}$) & $14.01\pm0.51$  \\ 
$a_\mathrm{c}$ (AU) & $0.0651\pm0.0024$  \\ 
$i_\mathrm{c}$ (deg) & $88.72_{-0.74}^{+0.82}$  \\ 
$e_\mathrm{c}$ & $0.18_{-0.14}^{+0.34}$  \\ 
$w_\mathrm{c}$ (deg) & $210\pm120$  \\ 
$b_\mathrm{tra;c}$ & $0.33_{-0.21}^{+0.25}$  \\ 
$T_\mathrm{tot;c}$ (h) & $2.913\pm0.095$  \\ 
$T_\mathrm{full;c}$ (h) & $2.754_{-0.094}^{+0.100}$  \\ 
$\rho_\mathrm{\star;c}$ (cgs) & $1.93\pm0.18$  \\ 
$T_\mathrm{eq;c}$ (K) & $932_{-16}^{+17}$  \\ 
$\delta_\mathrm{tr; c; TESS}$ (ppt) & $0.517_{-0.040}^{+0.036}$  \\
$P_\mathrm{c} / P_\mathrm{b}$ & $1.63461_{-0.00025}^{+0.00026}$  \\ 
$P_\mathrm{c} / P_\mathrm{d}$ & $0.437636\pm0.000052$  \\ 
$P_\mathrm{c} / P_\mathrm{e}$ & $0.316650\pm0.000046$  \\ 
        \end{tabular}
   \label{tabl:parainfefrst}
\end{table}

\begin{table}[!htbp]
	\centering
	\caption{Posterior of the inferred parameters for planets d and e and the host star.}
	\begin{tabular}{cc}

Property & Value \\ 
\hline 
\hline 
$R_\star/a_\mathrm{d}$ & $0.0365_{-0.0010}^{+0.0011}$  \\ 
$a_\mathrm{d}/R_\star$ & $27.39_{-0.82}^{+0.78}$  \\ 
$R_\mathrm{d}/a_\mathrm{d}$ & $0.001024_{-0.000046}^{+0.000048}$  \\ 
$R_\mathrm{d}$ ($\mathrm{R_{\oplus}}$) & $2.72\pm0.11$  \\ 
$R_\mathrm{d}$ ($\mathrm{R_{jup}}$) & $0.2423\pm0.0097$  \\ 
$a_\mathrm{d}$ ($\mathrm{R_{\odot}}$) & $24.31\pm0.87$  \\ 
$a_\mathrm{d}$ (AU) & $0.1131\pm0.0040$  \\ 
$i_\mathrm{d}$ (deg) & $89.22_{-0.38}^{+0.45}$  \\ 
$e_\mathrm{d}$ & $0.17_{-0.12}^{+0.30}$  \\ 
$w_\mathrm{d}$ (deg) & $190_{-130}^{+140}$  \\ 
$b_\mathrm{tra;d}$ & $0.35_{-0.21}^{+0.19}$  \\ 
$T_\mathrm{tot;d}$ (h) & $3.734_{-0.049}^{+0.066}$  \\ 
$T_\mathrm{full;d}$ (h) & $3.491_{-0.057}^{+0.061}$  \\ 
$\rho_\mathrm{\star;d}$ (cgs) & $1.93\pm0.17$  \\ 
$T_\mathrm{eq;d}$ (K) & $708_{-12}^{+13}$  \\ 
$\delta_\mathrm{tr; d; TESS}$ (ppt) & $0.889\pm0.053$  \\ 
$P_\mathrm{d} / P_\mathrm{b}$ & $3.73509\pm0.00053$  \\ 
$P_\mathrm{d} / P_\mathrm{c}$ & $2.28501\pm0.00027$  \\ 
$P_\mathrm{d} / P_\mathrm{e}$ & $0.723548_{-0.000097}^{+0.000090}$  \\ 
$R_\star/a_\mathrm{e}$ & $0.02948_{-0.00086}^{+0.00097}$  \\ 
$a_\mathrm{e}/R_\star$ & $33.9_{-1.1}^{+1.0}$  \\ 
$R_\mathrm{e}/a_\mathrm{e}$ & $0.000951_{-0.000043}^{+0.000049}$  \\ 
$R_\mathrm{e}$ ($\mathrm{R_{\oplus}}$) & $3.12_{-0.12}^{+0.13}$  \\ 
$R_\mathrm{e}$ ($\mathrm{R_{jup}}$) & $0.279_{-0.011}^{+0.012}$  \\ 
$a_\mathrm{e}$ ($\mathrm{R_{\odot}}$) & $30.1\pm1.1$  \\ 
$a_\mathrm{e}$ (AU) & $0.1400\pm0.0052$  \\ 
$i_\mathrm{e}$ (deg) & $89.32_{-0.30}^{+0.42}$  \\ 
$e_\mathrm{e}$ & $0.20_{-0.13}^{+0.30}$  \\ 
$w_\mathrm{e}$ (deg) & $170_{-130}^{+150}$  \\ 
$b_\mathrm{tra;e}$ & $0.36_{-0.23}^{+0.20}$  \\ 
$T_\mathrm{tot;e}$ (h) & $4.013_{-0.057}^{+0.080}$  \\ 
$T_\mathrm{full;e}$ (h) & $3.712_{-0.069}^{+0.063}$  \\ 
$\rho_\mathrm{\star;e}$ (cgs) & $1.92\pm0.18$  \\ 
$T_\mathrm{eq;e}$ (K) & $636_{-11}^{+12}$  \\ 
$\delta_\mathrm{tr; e; TESS}$ (ppt) & $1.175\pm0.069$  \\ 
$P_\mathrm{e} / P_\mathrm{b}$ & $5.16220\pm0.00084$ \\ 
$P_\mathrm{e} / P_\mathrm{c}$ & $3.15806\pm0.00046$ \\ 
$P_\mathrm{e} / P_\mathrm{d}$ & $1.38208_{-0.00017}^{+0.00019}$ \\ 
\hline
Limb darkening $u_\mathrm{1; TESS}$ & $0.40_{-0.24}^{+0.22}$ \\ 
Limb darkening $u_\mathrm{2; TESS}$ & $0.06_{-0.27}^{+0.36}$ \\
$\rho_\mathrm{\star; combined}$ (cgs) & $1.93\pm0.17$ \\  
        \end{tabular}
   \label{tabl:parainfeseco}
\end{table}

\bibliography{refr}

\suppressAffiliationsfalse
\allauthors

\end{document}